\documentclass[12pt]{article}

\usepackage{axodraw}
\usepackage{feynarts}

\makeatletter

\def\paragraph{\@startsection{paragraph}{4}{\z@}{+2.00ex plus
 +1ex minus +.2ex}{1.5ex plus .2ex}{\it\normalsize}}

\def\section{\@startsection {section}{1}{\z@}{+3.0ex plus +1ex minus
  +.2ex}{2.3ex plus .2ex}{\normalsize\bf\boldmath}}
\def\subsection{\@startsection{subsection}{2}{\z@}{+2.5ex plus +1ex
minus +.2ex}{1.5ex plus .2ex}{\normalsize\bf\boldmath}}
\def\subsubsection{\@startsection{subsubsection}{3}{\z@}{+3.25ex plus
 +1ex minus +.2ex}{1.5ex plus .2ex}{\normalsize\it}}

 \expandafter\ifx\csname mathrm\endcsname\relax\def\mathrm#1{{\rm #1}}\fi

\newcounter{saveeqn}

\@addtoreset{equation}{section}

\newcount\@tempcntc
\def\@citex[#1]#2{\if@filesw\immediate\write\@auxout{\string\citation{#2}}\fi
  \@tempcnta\z@\@tempcntb\m@ne\def\@citea{}\@cite{\@for\@citeb:=#2\do
    {\@ifundefined
       {b@\@citeb}{\@citeo\@tempcntb\m@ne\@citea
        \def\@citea{,\penalty\@m\ }{\bf ?}\@warning
       {Citation `\@citeb' on page \thepage \space undefined}}%
    {\setbox\z@\hbox{\global\@tempcntc0\csname
b@\@citeb\endcsname\relax}%
     \ifnum\@tempcntc=\z@ \@citeo\@tempcntb\m@ne
       \@citea\def\@citea{,\penalty\@m}
       \hbox{\csname b@\@citeb\endcsname}%
     \else
      \advance\@tempcntb\@ne
      \ifnum\@tempcntb=\@tempcntc
      \else\advance\@tempcntb\m@ne\@citeo
      \@tempcnta\@tempcntc\@tempcntb\@tempcntc\fi\fi}}\@citeo}{#1}}

\def\@citeo{\ifnum\@tempcnta>\@tempcntb\else\@citea
  \def\@citea{,\penalty\@m}%
  \ifnum\@tempcnta=\@tempcntb\the\@tempcnta\else
   {\advance\@tempcnta\@ne\ifnum\@tempcnta=\@tempcntb \else
\def\@citea{--}\fi
    \advance\@tempcnta\m@ne\the\@tempcnta\@citea\the\@tempcntb}\fi\fi}

\def\nl{\nonumber\\}

\newcommand{\lsim}
{\mathrel{\raisebox{-.3em}{$\stackrel{\displaystyle <}{\sim}$}}}
\newcommand{\gsim}
{\mathrel{\raisebox{-.3em}{$\stackrel{\displaystyle >}{\sim}$}}}
\def\asymp#1%
{\mathrel{\raisebox{-.4em}{$\widetilde{\scriptstyle #1}$}}}

\def\Nequal#1%
{\mathrel{\raisebox{-.5em}{$\stackrel{=}{\scriptstyle\rm#1}$}}}
\newcommand{\dsl}[1]{\not \hspace{-0.7mm}#1}
\def\dsl{\mathpalette\make@slash}
\def\make@slash#1#2{\setbox\z@\hbox{$#1#2$}%
  \hbox to 0pt{\hss$#1/$\hss\kern-\wd0}\box0}

\def\beq{\begin{equation}}
\def\eeq{\end{equation}}
\def\beqar{\begin{eqnarray}}
\def\eeqar{\end{eqnarray}}
\def\barr#1{\begin{array}{#1}}
\def\earr{\end{array}}
\def\bfi{\begin{figure}}
\def\efi{\end{figure}}
\def\btab{\begin{table}}
\def\etab{\end{table}}
\def\bce{\begin{center}}
\def\ece{\end{center}}
\def\nn{\nonumber}

\def\text{\textstyle}

\def\arraystretch{1.2}

\def\al{\alpha}
\def\be{\beta}
\def\Ga{\Gamma}
\def\ga{\gamma}
\def\de{\delta}
\def\De{\Delta}

\def\la{\lambda}

\def\si{\sigma}

\def\refeq#1{\mbox{(\ref{#1})}}

\def\reffi#1{\mbox{Figure~\ref{#1}}}
\def\reffis#1{\mbox{Figures~\ref{#1}}}
\def\refta#1{\mbox{Table~\ref{#1}}}

\def\refse#1{\mbox{Section~\ref{#1}}}

\def\citere#1{\mbox{Ref.~\cite{#1}}}
\def\citeres#1{\mbox{Refs.~\cite{#1}}}

\newcommand{\GeV}{\unskip\,\mathrm{GeV}}
\newcommand{\MeV}{\unskip\,\mathrm{MeV}}

\newcommand{\ri}{{\mathrm{i}}}
\newcommand{\rd}{{\mathrm{d}}}

\newcommand{\rT}{{\mathrm{T}}}

\newcommand{\Ord}{\mathswitch{{\cal{O}}}}
\newcommand{\Oa}{\mathswitch{{\cal{O}}(\alpha)}}

\newcommand{\Oaaa}{\mathswitch{{\cal{O}}(\alpha^3)}}

\newcommand{\M}{{\cal{M}}}

\def\mathswitchr#1{\relax\ifmmode{\mathrm{#1}}\else$\mathrm{#1}$\fi}
\newcommand{\Pf}{\mathswitch  f}

\newcommand{\PV}{V}
\newcommand{\PW}{\mathswitchr W}

\newcommand{\PZ}{\mathswitchr Z}

\newcommand{\Pg}{\mathswitchr g}

\newcommand{\PH}{\mathswitchr H}
\newcommand{\Pe}{\mathswitchr e}
\newcommand{\Pne}{\mathswitch \nu_{\mathrm{e}}}

\newcommand{\Pnmu}{\mathswitch \nu_{\mu}}
\newcommand{\Pnmubar}{\mathswitch \bar\nu_{\mu}}
\newcommand{\Pd}{\mathswitchr d}

\newcommand{\Pu}{\mathswitchr u}

\newcommand{\Ps}{\mathswitchr s}

\newcommand{\Pc}{\mathswitchr c}

\newcommand{\Pb}{\mathswitchr b}

\newcommand{\Pt}{\mathswitchr t}
\newcommand{\Ptbar}{\mathswitchr{\bar t}}
\newcommand{\Pep}{\mathswitchr {e^+}}
\newcommand{\Pem}{\mathswitchr {e^-}}

\newcommand{\Pmup}{\mathswitchr {\mu^+}}
\newcommand{\Pmum}{\mathswitchr {\mu^-}}

\def\mathswitch#1{\relax\ifmmode#1\else$#1$\fi}

\newcommand{\MW}{\mathswitch {M_\PW}}

\newcommand{\MZ}{\mathswitch {M_\PZ}}
\newcommand{\MH}{\mathswitch {M_\PH}}
\newcommand{\Me}{\mathswitch {m_\Pe}}
\newcommand{\Mmy}{\mathswitch {m_\mu}}
\newcommand{\Mta}{\mathswitch {m_\tau}}
\newcommand{\Md}{\mathswitch {m_\Pd}}
\newcommand{\Mu}{\mathswitch {m_\Pu}}
\newcommand{\Ms}{\mathswitch {m_\Ps}}
\newcommand{\Mc}{\mathswitch {m_\Pc}}
\newcommand{\Mb}{\mathswitch {m_\Pb}}
\newcommand{\Mt}{\mathswitch {m_\Pt}}
\newcommand{\GW}{\Gamma_{\PW}}

\newcommand{\GZ}{\Gamma_{\PZ}}

\newcommand{\Gt}{\Gamma_{\Pt}}

\newcommand{\rw}{\mathswitchr w}
\newcommand{\sw}{\mathswitch {s_\rw}}
\newcommand{\cw}{\mathswitch {c_\rw}}

\newcommand{\Qf}{\mathswitch {Q_\Pf}}

\newcommand{\GF}{\mathswitch {G_\mu}}

\def\solid{\raise.9mm\hbox{\protect\rule{1.1cm}{.2mm}}}
\def\dash{\raise.9mm\hbox{\protect\rule{2mm}{.2mm}}\hspace*{1mm}}

\def\ie{i.e.\ }

\newcommand{\QCD}{{\mathrm{QCD}}}

\newcommand{\LEP}{{\mathrm{LEP}}}

\newcommand{\IBA}{{\mathrm{IBA}}}
\newcommand{\Coul}{{\mathrm{Coul}}}

\newcommand{\WW}{{\mathrm{WW}}}
\newcommand{\ZZ}{{\mathrm{ZZ}}}

\newcommand{\NWA}{\mathswitch{\mathrm{NWA}}}

\newcommand{\LL}{\mathswitch{\mathrm{LL}}}
\newcommand{\LLFSR}{\mathswitch{\mathrm{LLFSR}}}

\def\Li{\mathop{\mathrm{Li}_2}\nolimits}

\def\Re{\mathop{\mathrm{Re}}\nolimits}
\def\Im{\mathop{\mathrm{Im}}\nolimits}
\def\sgn{\mathop{\mathrm{sgn}}\nolimits}
\def\lra{\mathop{\mathrm{\leftrightarrow}}\nolimits}

\newcommand{\HDECAY}{{\sc Hdecay}}

\def\draftdate{\relax}
\def\mda{\relax}
\def\mua{\relax}
\def\mla{\relax}
\def\Mda{\relax}
\def\Mua{\relax}
\def\Mla{\relax}
\def\draft{
\def\thtystars{******************************}
\def\sixtystars{\thtystars\thtystars}
\typeout{}
\typeout{\sixtystars**}
\typeout{* Draft mode!
         For final version remove \protect\draft\space in source file *}
\typeout{\sixtystars**}
\typeout{}
\def\draftdate{\today}
\def\mua{\marginpar[\boldmath\hfil$\uparrow$]%
                   {\boldmath$\uparrow$\hfil}%
                    \typeout{marginpar: $\uparrow$}\ignorespaces}
\def\mda{\marginpar[\boldmath\hfil$\downarrow$]%
                   {\boldmath$\downarrow$\hfil}%
                    \typeout{marginpar: $\downarrow$}\ignorespaces}
\def\mla{\marginpar[\boldmath\hfil$\rightarrow$]%
                   {\boldmath$\leftarrow $\hfil}%
                    \typeout{marginpar: $\lra$}\ignorespaces}
\def\Mua{\marginpar[\boldmath\hfil$\Uparrow$]%
                   {\boldmath$\Uparrow$\hfil}%
                    \typeout{marginpar: $\uparrow$}\ignorespaces}
\def\Mda{\marginpar[\boldmath\hfil$\Downarrow$]%
                   {\boldmath$\Downarrow$\hfil}%
                    \typeout{marginpar: $\downarrow$}\ignorespaces}
\def\Mla{\marginpar[\boldmath\hfil$\Rightarrow$]%
                   {\boldmath$\Leftarrow $\hfil}%
                    \typeout{marginpar: $\lra$}\ignorespaces}
\overfullrule 5pt
\oddsidemargin -15mm
\marginparwidth 29mm
}

\def\stars{\strut\leaders\hbox{*}\hfill\strut}
\def\starline{\hfil\strut\hfil\hbox to \textwidth {\stars}\hfil}

\begin{document}
\thispagestyle{empty}
\def\thefootnote{\fnsymbol{footnote}}
\setcounter{footnote}{1}
\null
\draftdate\hfill MPP-2005-24\\
\strut\hfill PSI-PR-06-05\\
\strut\hfill hep-ph/0604011\\
\vspace{1.5cm}
\begin{center}
{\Large \bf\boldmath
Precise predictions for the Higgs-boson decay 
\\[.5em]
$\PH\to\PW\PW/\PZ\PZ\to4\,$leptons
\par} 
\vspace{1cm}
{\large
{\sc A.\ Bredenstein$^1$, A.\ Denner$^2$, S.\ Dittmaier$^1$ 
and M.M.\ Weber$^3$} } \\[1cm]
$^1$ {\it Max-Planck-Institut f\"ur Physik
(Werner-Heisenberg-Institut), \\
D-80805 M\"unchen, Germany}
\\[0.5cm]
$^2$ {\it Paul Scherrer Institut, W\"urenlingen und Villigen,
\\
CH-5232 Villigen PSI, Switzerland} \\[0.5cm]
$^3$ {\it Fachbereich Physik, Bergische Universit\"at Wuppertal,
\\
D-42097 Wuppertal, Germany}
\par \vskip 2em
\end{center}\par
\vfill {\bf Abstract:} \par The decay of the Standard Model Higgs
boson into four leptons via a virtual W-boson or Z-boson pair is one
of the most important decay modes in the Higgs-boson search at the
LHC. We present the complete electroweak radiative corrections of
${\cal O}(\alpha)$ to these processes, including improvements beyond
${\cal O}(\alpha)$ originating from heavy-Higgs effects and
final-state radiation.  The intermediate W- and Z-boson resonances are
described (without any expansion or on-shell approximation) by
consistently employing complex mass parameters for the gauge bosons
(complex-mass scheme).  The corrections to partial decay widths
typically amount to some per cent and increase with growing Higgs mass
$\MH$, reaching about 8\% at $\MH\sim500\GeV$.  For not too large
Higgs masses ($\MH\lsim400\GeV$) the corrections to the partial decay
widths can be reproduced within $\lsim2\%$ by simple approximations.
For angular distributions the corrections are somewhat larger and
distort the shapes.  For invariant-mass distributions of fermion pairs
they can reach several tens of per cent depending on the treatment of
photon radiation.  The discussed corrections have been implemented in
a Monte Carlo
event generator called {\sc Prophecy4f}.%
\footnote{The computer code can be obtained from the authors upon request.}%
\par
\vskip .5cm
\noindent
March 2006 
\null
\setcounter{page}{0}
\clearpage
\def\thefootnote{\arabic{footnote}}
\setcounter{footnote}{0}

\section{Introduction}

The primary task of the LHC will be the detection and the study of the
Higgs boson. If it is heavier than $140\GeV$ and behaves as predicted
by the Standard Model (SM), it decays predominantly into gauge-boson
pairs and subsequently into four light fermions.  From a Higgs-boson
mass $\MH$ of about $130\GeV$ up to the Z-boson-pair threshold $2\MZ$,
the decay signature $\PH(\to\PW\PW)\to2\,$leptons + missing
$p_{\mathrm{T}}$ \cite{Glover:1988fn,Dittmar:1996ss} has the highest
discovery potential for the Higgs boson at the LHC \cite{Asai:2004ws}.
For higher Higgs masses, the leading role is taken over by the
``gold-plated'' channel $\PH\to\PZ\PZ\to4\,$leptons, which will allow
for the most accurate measurement of $\MH$ above $130\GeV$
\cite{Zivkovic:2004sv}.  More details and recent developments
concerning Higgs studies at the LHC can be found in the literature
\cite{atlas-cms-tdrs,Assamagan:2004mu,SMH-LH2005}.  At a future
$\Pe^+\Pe^-$ linear collider
\cite{Aguilar-Saavedra:2001rg,Abe:2001wn,Abe:2001gc}, the decays
$\PH\to4f$ will enable measurements of the $\PH\to\PW\PW/\PZ\PZ$
branching ratios at the level of a few to 10\% \cite{Meyer:2004ha}.

A kinematical reconstruction of the Higgs boson and of the virtual W
and Z~bosons requires the study of distributions defined from the
kinematics of the decay fermions.  Thereby, it is important to include
radiative corrections, in particular real photon radiation. In
addition, the verification of the spin and of the CP properties of the
Higgs boson relies on the study of angular, energy, and invariant-mass
distributions \cite{Nelson:1986ki,Choi:2002jk}.  In particular, the
sensitivity of the angle between the two Z-decay planes in
$\PH\to\PZ\PZ\to4\,$leptons has been frequently emphasized in the
literature.  As a consequence a Monte Carlo generator for
$\PH\to\PW\PW/\PZ\PZ\to4\,$fermions including all relevant corrections
is needed.

The theoretical description of the decays of a SM Higgs boson into W-
or Z-boson pairs started with lowest-order formulas for the partial
decay widths.  The first calculations \cite{Pocsik:1980ta} that
include off-shell effects of the gauge bosons made the approximation
that one of the W or Z~bosons was still on shell, an approximation
that turns out to be not sufficient.  Later calculations
\cite{Cahn:1988ru} dealt with the situation of two intermediate
off-shell gauge bosons.  The various approaches are compared, e.g., in
\citere{Djouadi:2005gi}.  We note that the program
\HDECAY~\cite{Djouadi:1997yw}, which is frequently used in practice,
calculates the partial decay widths for $\PH\to\PW\PW/\PZ\PZ$ with on-
or off-shell gauge bosons depending on $\MH$.  Distributions of the
decay fermions have been considered in
\citeres{Nelson:1986ki,Choi:2002jk}, but still in lowest order of
perturbation theory.

In the past the electroweak ${\cal O}(\alpha)$ corrections to decays
into gauge bosons, $\PH\to\PW\PW/\PZ\PZ$, were known
\cite{Fleischer:1980ub,Kniehl:1991xe} only in narrow-width
approximation (NWA), i.e.\ for on-shell W and Z~bosons.  In this case,
also leading two-loop corrections enhanced by powers of the top-quark
mass \cite{Kniehl:1995tn,Kniehl:1995at} or of the Higgs-boson mass
\cite{Ghinculov:1995bz,Frink:1996sv} have been calculated.  However,
near and below the gauge-boson-pair thresholds the NWA is not
applicable, so that only the lowest-order results exist in this $\MH$
range.  Recently electroweak corrections to the processes
$\PH\to\PW\PW/\PZ\PZ\to4f$ with off-shell gauge bosons have been
considered. Progress on a calculation of the electromagnetic
corrections to $\PH\to\PZ\PZ\to4\,$leptons has been reported at the
RADCOR05 conference by Carloni Calame \cite{CarloniCalame:2006vr}.
There we have also presented first results of our calculation of the
complete ${\cal O}(\alpha)$ corrections to the general
$\PH\to4\,$leptons processes \cite{axel-talk}.

In this paper we describe the details of our calculation of the ${\cal
  O}(\alpha)$ corrections and of the included improvements beyond this
order.  The involved Feynman diagrams are closely related to the ones
of the production process $\Pep\Pem\to\nu\bar\nu\PH$, whose
electroweak ${\cal O}(\alpha)$ corrections have been evaluated in
\citeres{Belanger:2002me,Denner:2003yg}.  Therefore, we proceed in the
algebraic reduction of the one-loop diagrams as described in
\citere{Denner:2003yg}.  On the other hand, the resonance structure of
the decays $\PH\to\PW\PW/\PZ\PZ\to4f$ is practically the same as in
$\Pep\Pem\to\PW\PW\to4f$, which was treated at the one-loop level in
\citere{Denner:2005es}.  Thus, we apply the ``complex-mass scheme''
\cite{Denner:2005es,Denner:1999gp}, where gauge-boson masses are
consistently treated as complex quantities.  This procedure fully
maintains gauge invariance at the price of having complex gauge-boson
masses everywhere, i.e.\ also in couplings and loop integrals. For a
numerically stable evaluation of the latter we employ the methods
described in \citeres{Denner:2002ii,Denner:2005nn}.  The combination
of virtual and real photon corrections is performed in the dipole
subtraction approach \cite{Dittmaier:2000mb,Bredenstein:2005zk} and
checked by the alternative of phase-space slicing.  The whole
calculation has been implemented in a Monte Carlo generator called
{\sc Prophecy4f}.

The paper is organized as follows.  In \refse{se:conv-lo} we fix our
conventions and give explicit results for the tree-level amplitudes of
the processes $\PH\to\PW\PW/\PZ\PZ\to4f$.  Section~\ref{se:virt}
contains a description of our calculation of the virtual one-loop
corrections; the real photon corrections are considered in
\refse{se:real}.  Some details on the employed Monte Carlo techniques
are given in \refse{se:MC}.  In \refse{se:IBA} we construct an
``improved Born approximation'' (IBA) which approximates our
state-of-the-art prediction for partial decay widths within $\lsim2\%$
for not too large Higgs masses, $\MH\lsim400\GeV$.  Our numerical
results are discussed in \refse{se:numerics}, comprising partial decay
widths of several representative $\PH\to\PW\PW/\PZ\PZ\to4l$ channels
as well as differential cross sections for selected channels in
invariant masses of lepton pairs and in various angles.  A comparison
with results obtained from \HDECAY\ for the partial widths is also
performed there.  Section~\ref{se:concl} contains our conclusions.

\section{Conventions and lowest-order results}
\label{se:conv-lo}

We consider the lowest-order processes 
\beq\label{process-H4f}
\PH(p)
 \;\longrightarrow\;
f_1(k_1,\si_1) + \bar f_2(k_2,\si_2) + f_3(k_3,\si_3) + \bar
f_4(k_4,\si_4),  
\label{eq:h4f}
\eeq
where the momenta and helicities of the external particles are
indicated in parentheses. The helicities take the values
$\sigma_i=\pm1/2$, but we often use only the sign to indicate the
helicity. The masses of the external fermions are neglected whenever 
possible; they are only taken into account in the mass-singular
logarithms originating from collinear final-state radiation (FSR).
The matrix elements can be constructed from the generic diagram shown in
\reffi{fi:H4f-born-diag}.
\bfi
\begin{center}
\setlength{\unitlength}{1pt}
\begin{picture}(190,100)(-20,0)
\DashLine(15,50)(60,50){3}
\Photon(60,50)(90,20){-2}{5}
\Photon(60,50)(90,80){2}{5}
\Vertex(60,50){2.0}
\Vertex(90,80){2.0}
\Vertex(90,20){2.0}
\ArrowLine(90,80)(120, 95)
\ArrowLine(120,65)(90,80)
\ArrowLine(120, 5)( 90,20)
\ArrowLine( 90,20)(120,35)
\put(-20,47){$\PH(p)$}
\put(62,70){$V$}
\put(62,18){$V$}
\put(125,90){$f_a(k_a,\si_a)$}
\put(125,65){$\bar f_b(k_b,\si_b)$}
\put(125,30){$f_c(k_c,\si_c)$}
\put(125,5){$\bar f_d(k_d,\si_d)$}
\end{picture}
\end{center}
\caption{Generic lowest-order diagram for $\PH\to 4f$ where $V=\PW,\PZ$.}
\label{fi:H4f-born-diag}
\efi
\newcommand{\cmhs}{\mu^2_\PH}
\newcommand{\cmt}{\mu_\Pt}
\newcommand{\cmvs}{\mu^2_V}
\newcommand{\cmws}{\mu^2_\PW}
\newcommand{\cmzs}{\mu^2_\PZ}
\newcommand{\cmv}{\mu_V}
\newcommand{\cmw}{\mu_\PW}
\newcommand{\cmz}{\mu_\PZ}
\newcommand{\csw}{\mathswitch {s_\rw}}
\newcommand{\ccw}{\mathswitch {c_\rw}}
\newcommand{\cZ}{\mathcal{Z}}

Using the conventions of \citeres{Denner:1993kt,Denner:1994xt} we denote 
the relevant couplings in the following by
\beqar
g_{\gamma ff}^\pm &=& -\Qf, \qquad
g_{\PZ ff}^+ = -\frac{\sw}{\cw}\Qf, \qquad
g_{\PZ ff}^- = -\frac{\sw}{\cw}\Qf + \frac{I^3_{\rw,f}}{\cw\sw},
\nn\\
g_{\PW ff'}^- &=& \frac{1}{\sqrt{2}\sw}, \qquad
g_{\PW ff'}^+ = 0,
\nl
g_{\PH\PZ\PZ}&=&\frac{\cmw}{\cw^2\sw},\qquad g_{\PH\PW\PW}=\frac{\cmw}{\sw},
\label{eq:couplings}
\eeqar
where $\Qf$ is the relative charge of the fermion $f$, and
$I^3_{\rw,f}=\pm{1}/{2}$ the third component of the weak isospin of
the left-handed part of the fermion field $f$.  The CKM matrix has
been consistently set to the unit matrix, which has no sizeable
effects on our results.  In \refeq{eq:couplings} we have already
indicated that we use complex gauge-boson masses $\cmv$ everywhere,
\beqar\label{eq:complex-masses}
\cmvs&=& M_V^2 - \ri M_V\Ga_V, \qquad V=\PW,\PZ,
\eeqar
where $M_V$ and $\Ga_V$ denote the real pole-mass and
width parameters. Accordingly the sine and cosine of
the weak mixing angle are fixed by
\beq\label{eq:defcw}
\cw^2 = 1-\sw^2 = \frac{\cmw^2}{\cmz^2},
\eeq
i.e.\ $\cw$ and $\sw$ are complex quantities.
More details about the complex-mass scheme are given in
\refse{se:cms}.

In order to express the amplitudes in a compact way, we use the
Weyl--van der Waerden (WvdW) spinor technique as formulated in
\citere{Dittmaier:1998nn}. The spinor products $\langle\dots\rangle$
are defined by
\beq
\langle pq\rangle=\epsilon^{AB}p_A q_B
=2\sqrt{p_0 q_0} \,\Biggl[
{\mathrm{e}}^{-\ri\phi_p}\cos\frac{\theta_p}{2}\sin\frac{\theta_q}{2}
-{\mathrm{e}}^{-\ri\phi_q}\cos\frac{\theta_q}{2}\sin\frac{\theta_p}{2}
\Biggr],
\eeq
where $p_A$, $q_A$ are the associated momentum spinors for the light-like
momenta
\beqar
p^\mu&=&p_0(1,\sin\theta_p\cos\phi_p,\sin\theta_p\sin\phi_p,\cos\theta_p),\nl
q^\mu&=&q_0(1,\sin\theta_q\cos\phi_q,\sin\theta_q\sin\phi_q,\cos\theta_q).
\eeqar

In the notation of \citere{Dittmaier:1998nn} the generic lowest-order 
amplitude reads
\beq\label{eq:MHffff}
\M_0^{VV,\si_a\si_b\si_c\si_d}(k_a,k_b,k_c,k_d) =
2e^3g^{\si_a}_{Vf_af_b}g^{\si_c}_{Vf_cf_d}g_{\PH VV} \,
\de_{\si_a,-\si_b}\de_{\si_c,-\si_d}
A_{\si_a\si_c}^{VV}(k_a,k_b,k_c,k_d),
\eeq
or more specifically for the case of Z-mediated and W-mediated
decays
\beqar\label{eq:MHffff-spec}
\M_0^{\ZZ,\si_a\si_b\si_c\si_d}(k_a,k_b,k_c,k_d) &=&
\frac{2e^3g^{\si_a}_{\PZ f_af_b}g^{\si_c}_{\PZ f_cf_d}\cmw}{\cw^2\sw} \,
\de_{\si_a,-\si_b}\de_{\si_c,-\si_d}\,
A_{\si_a\si_c}^{\ZZ}(k_a,k_b,k_c,k_d),
\nl
\M_0^{\WW,\si_a\si_b\si_c\si_d}(k_a,k_b,k_c,k_d) &=&
 \frac{e^3\cmw}{\sw^3} \,
\de_{\si_a,-}\de_{\si_b,+}\de_{\si_c,-}\de_{\si_d,+}\,
A_{--}^{\WW}(k_a,k_b,k_c,k_d).
\eeqar
The auxiliary functions are given by
\beqar\label{eq:Mhffffaux}
A_{--}^{VV}(k_a,k_b,k_c,k_d) &=& 
\frac{\langle k_b k_d\rangle^*\langle k_a k_c\rangle}
{[(k_a+k_b)^2-\cmv^2] [(k_c+k_d)^2-\cmv^2]},
\nl
A_{+-}^{VV}(k_a,k_b,k_c,k_d) &=& A_{--}^{VV}(k_b,k_a,k_c,k_d),\nl
A_{-+}^{VV}(k_a,k_b,k_c,k_d) &=& A_{--}^{VV}(k_a,k_b,k_d,k_c),\nl
A_{++}^{VV}(k_a,k_b,k_c,k_d) &=& A_{--}^{VV}(k_b,k_a,k_d,k_c),
\eeqar
and obey the relations
\beqar\label{eq:MHffff-rel}
A_{-\si_a,-\si_c}^{VV}(k_a,k_b,k_c,k_d) &=& 
\Bigl(A_{\si_a\si_c}^{VV}(k_a,k_b,k_c,k_d)\Bigr)^*
\Bigr|_{\cmv\to \cmv^*},\nl
A_{-\si_a,\si_c}^{VV}(k_a,k_b,k_c,k_d) &=& 
A_{\si_a\si_c}^{VV}(k_b,k_a,k_c,k_d), \nl
A_{\si_a,-\si_c}^{VV}(k_a,k_b,k_c,k_d) &=& 
A_{\si_a\si_c}^{VV}(k_a,k_b,k_d,k_c), \nl
A_{\si_a\si_c}^{VV}(k_a,k_b,k_c,k_d) &=& 
\Bigl(A_{\si_a\si_c}^{VV}(k_b,k_a,k_d,k_c)\Bigr)^*
\Bigr|_{\cmv\to \cmv^*},\nl
A_{\si_a\si_c}^{VV}(k_a,k_b,k_c,k_d) &=& 
A_{\si_c\si_a}^{VV}(k_c,k_d,k_a,k_b).
\eeqar
The relations between the $A^{\dots}_{\dots}$ functions that differ in
all helicities result from a P transformation. Those where only one
fermion helicity is reversed are related to C symmetry.  The last but
one is due to CP symmetry, and the last one results from a symmetry
under the exchange of the two fermion pairs.  The replacements
$\cmv\to \cmv^*$ in \refeq{eq:MHffff-rel} ensure that the vector-boson
masses remain unaffected by complex conjugation.

From the generic matrix element
$\M_0^{VV,\si_a\si_b\si_c\si_d}(k_a,k_b,k_c,k_d)$ the matrix elements
for the specific processes can be constructed as follows.  To write
down the explicit matrix elements for the different final states, we
denote different fermions ($f\ne F$) by $f$ and $F$, and their
weak-isospin partners by $f'$ and $F'$, respectively:
\begin{itemize}
\item $\PH\to f\bar f F\bar F$:
\beqar
\M_0^{\si_1\si_2\si_3\si_4}(k_1,k_2,k_3,k_4)&=&
\M_0^{\ZZ,\si_1\si_2\si_3\si_4}(k_1,k_2,k_3,k_4),
\label{eq:zz}
\eeqar
\item $\PH\to f\bar f' F\bar F'$:
\beqar
\M_0^{\si_1\si_2\si_3\si_4}(k_1,k_2,k_3,k_4)
&=&\M_0^{\WW,\si_1\si_2\si_3\si_4}(k_1,k_2,k_3,k_4),
\label{eq:ww}
\eeqar
\item $\PH\to f\bar f f\bar f$:
\beqar
\M_0^{\si_1\si_2\si_3\si_4}(k_1,k_2,k_3,k_4)&=&
\M_0^{\ZZ,\si_1\si_2\si_3\si_4}(k_1,k_2,k_3,k_4)
\nl&&{}
-\M_0^{\ZZ,\si_1\si_4\si_3\si_2}(k_1,k_4,k_3,k_2),
\label{eq:zzsym}
\eeqar
\item $\PH\to f\bar f f'\bar f'$:
\beqar
\M_0^{\si_1\si_2\si_3\si_4}(k_1,k_2,k_3,k_4)
&=&\M_0^{\ZZ,\si_1\si_2\si_3\si_4}(k_1,k_2,k_3,k_4)
\nl&&{}
-\M_0^{\WW,\si_1\si_4\si_3\si_2}(k_1,k_4,k_3,k_2).
\label{eq:mixed}
\eeqar
\end{itemize}
The relative signs between contributions of the basic subamplitudes to
the full matrix elements account for the sign changes resulting from
interchanging external fermion lines.

The matrix elements of \refeq{eq:zz} and \refeq{eq:ww} can be extended
to the case of semi-leptonic or hadronic final states by simply
multiplying the squared matrix element by a colour factor 3 or 9,
respectively.  On the other hand, care has to be taken in the cases of
\refeq{eq:zzsym} and \refeq{eq:mixed} for hadronic final states
(semi-leptonic final states do not exist) owing to the non-trivial
colour interferences.  Summing over the colour degrees of freedom, we
have
\begin{itemize}
\item $\PH\to q\bar q q\bar q$:
\beqar
\lefteqn{\left|\M_0^{\si_1\si_2\si_3\si_4}(k_1,k_2,k_3,k_4)\right|^2 =}
\qquad
\nl&&
9\left|\M_0^{\ZZ,\si_1\si_2\si_3\si_4}(k_1,k_2,k_3,k_4)\right|^2 
+9\left|\M_0^{\ZZ,\si_1\si_4\si_3\si_2}(k_1,k_4,k_3,k_2)\right|^2
\nl&&{}
-6\Re\left\{\M_0^{\ZZ,\si_1\si_2\si_3\si_4}(k_1,k_2,k_3,k_4)
\left(\M_0^{\ZZ,\si_1\si_4\si_3\si_2}(k_1,k_4,k_3,k_2)\right)^*\right\},
\hspace*{3em}
\label{eq:hadzzsym}
\eeqar
\item $\PH\to q\bar q q'\bar q'$:
\beqar
\lefteqn{\left|\M_0^{\si_1\si_2\si_3\si_4}(k_1,k_2,k_3,k_4)\right|^2 =}
\qquad
\nl&&
9\left|\M_0^{\ZZ,\si_1\si_2\si_3\si_4}(k_1,k_2,k_3,k_4)\right|^2
+9\left|\M_0^{\WW,\si_1\si_4\si_3\si_2}(k_1,k_4,k_3,k_2)\right|^2
\nl&&
{}-6\Re\left\{\M_0^{\ZZ,\si_1\si_2\si_3\si_4}(k_1,k_2,k_3,k_4)
\left(\M_0^{\WW,\si_1\si_4\si_3\si_2}(k_1,k_4,k_3,k_2)\right)^*\right\}.
\hspace*{3em}
\label{eq:hadmixed}
\eeqar
\end{itemize}
Here $q$ denotes any quark of the first two generations and $q'$ its
weak-isospin partner.

Having constructed the matrix elements, we can write the lowest-order 
decay width $\Ga_0$ as
\beq
\int\rd\Ga_0 = \frac{1}{2\MH} \int \rd\Phi_0 \,
 \sum_{\si_1,\si_2,\si_3,\si_4=\pm\frac{1}{2}}  
|\M^{\si_1\si_2\si_3\si_4}_0|^2,
\label{eq:hbcs}
\eeq
where the phase-space integral is defined by
\beq
\int \rd\Phi_0 =
\left( \prod_{i=1}^4 \int\frac{\rd^3 {\bf k}_i}{(2\pi)^3 2k_i^0} \right)\,
(2\pi)^4 \delta^{(4)}\Biggl(p-\sum_{j=1}^4 k_j\Biggr).
\label{eq:dPS}
\eeq
For the case $\PH\to f\bar f f\bar f$, which involves two pairs of
identical particles in the final state, we implicitly include a factor
$1/4$ in the phase-space integral, without making this factor
explicit in the formulas.

\section{Virtual corrections}
\label{se:virt}

\subsection{Survey of one-loop diagrams}

\begin{figure}
\centerline{\footnotesize  \input{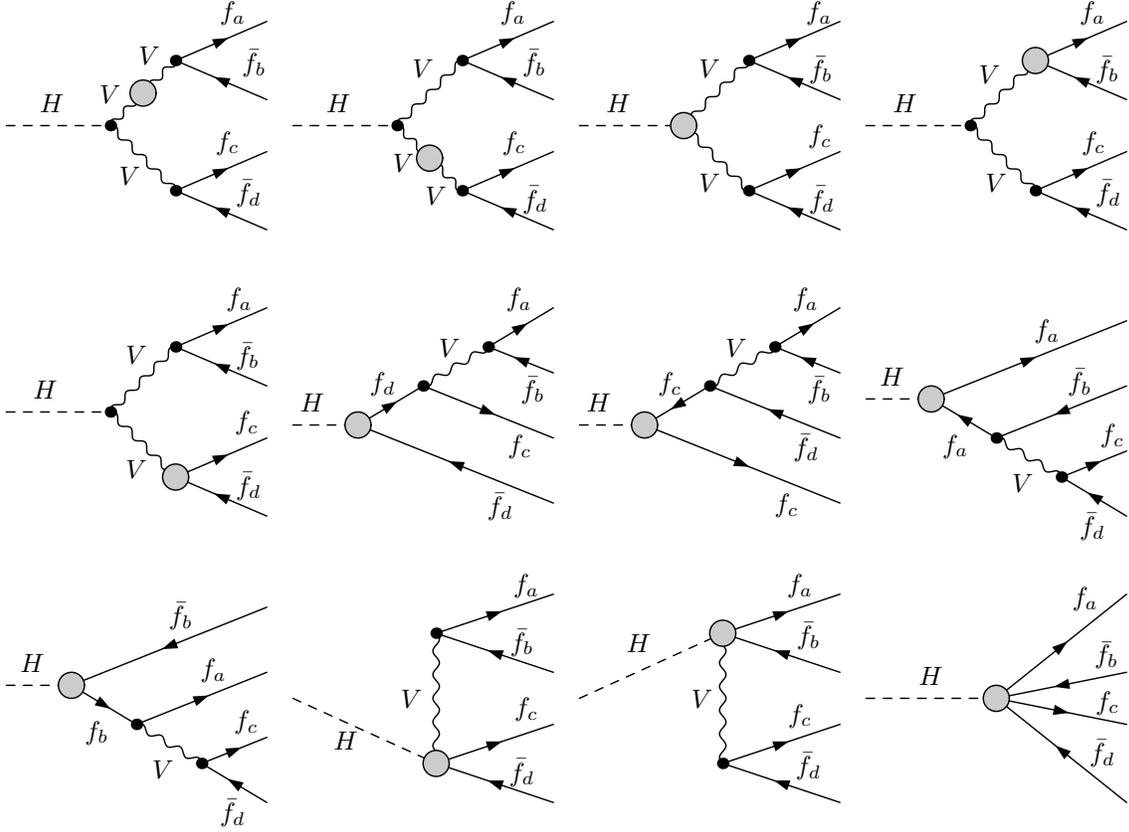}}
\vspace{-2.0em}
\caption{Generic contributions of different vertex functions to 
$\PH\to\PW\PW/\PZ\PZ\to4f$, where the blobs stand for one-particle-irreducible
one-loop vertex functions.}
\label{fi:gendiagrams}
\end{figure}

The virtual corrections receive contributions from self-energy,
vertex, box, and pentagon diagrams. The structural diagrams
containing the generic contributions of vertex functions are
summarized in \reffi{fi:gendiagrams}.  
Here and in the following we omit all diagrams that vanish in the
limit of vanishing external fermion masses from the beginning.  For
charged-current processes the generic field $V$ stands for the W-boson
field, for neutral-current processes we have $V=Z,\gamma$, where the
photon is absent in couplings to the Higgs boson.  The generic
diagrams cover all structures relevant for electroweak corrections to
arbitrary four-fermion final states, including quarks.  Note, however,
that some four-quark final states receive corrections from diagrams
with intermediate gluons on tree-like lines (quark-loop-induced
$\PH\Pg\Pg$ vertex).  Possible QCD corrections for quarks in the final
state will not be considered in the following lists of diagrams.

The pentagon diagrams are shown in \reffis{fi:pent_nc} and
\ref{fi:pent_cc}, respectively.
\begin{figure}
\centerline{\footnotesize  \input{paper-pent-nc}}
\vspace{-2.0em}
\caption{Pentagon diagrams for $\PH\to\PZ\PZ\to f\bar f F\bar F$,
  where $f$ and $F$ are different fermions with respective
  weak-isospin partners $f'$ and $F'$. The fermion arrows in the
  diagrams involving W~bosons apply to the case where both $f$ and $F$ have
  weak isospin $I^3_{\rw,f}= +1/2$; for fermions with $I^3_{\rw,f}=
  -1/2$ the corresponding fermion arrows have to be reversed.}
\label{fi:pent_nc}
%
\vspace*{1em}
\centerline{\footnotesize  \input{paper-pent-cc}}
\vspace{-2.0em}
\caption{Pentagon diagrams for $\PH\to\PW\PW\to f\bar f' F\bar F'$,
  where $f$ and $F$ are different fermions with respective
  weak-isospin partners $f'$ and $F'$. The fermion arrows apply to the case 
  where $f$ and $F$ have weak isospin $I^3_{\rw,f}= +1/2$ and
  $I^3_{\rw,F}= -1/2$, respectively; for fermions with opposite weak
  isospin the corresponding fermion arrows have to be reversed.}
\label{fi:pent_cc}
\end{figure}
The specific subdiagrams of loop-induced 4-point functions have been
shown in \citere{Denner:2003yg}, where the process class
$\Pep\Pem\to\nu\bar\nu\PH$ was analyzed at one loop. They involve
4-point vertex functions of the type $\nu_l\bar{\nu}_l\PZ\PH$,
$\nu_l\bar{\nu}_l\gamma\PH$, $l^-l^+\PZ\PH$, $l^-l^+\gamma\PH$, and
$l^\mp\mathord{\stackrel{{\scriptscriptstyle(}-{\scriptscriptstyle)}}{\nu}}_l\PW^\pm\PH$
with $l=\Pe,\mu,\tau$ denoting any charged lepton.  The diagrams for
the $l^-l^+\gamma\PH$ vertex function can be obtained from those for
the $l^-l^+\PZ\PH$ vertex function by replacing the external \PZ~boson
by a photon and omitting the diagram where the photon couples to
neutrinos.  The 3-point loop insertions in the $\PH\nu_l\bar{\nu}_l$,
$\PH l^-l^+$, $\PH\PW\PW$, $\PH\PZ\PZ$, and $\PH\PZ\ga$ vertices have
also been listed in \citere{Denner:2003yg}; the one-loop diagrams for
the $\PH\gamma\gamma$ vertex follow from the $\PH\PZ\PZ$ or
$\PH\PZ\gamma$ case by obvious substitutions and omissions.  Most of
the diagrams for the self-energies and the $\nu_l \bar{\nu}_l\PZ$,
$l^-l^+\PZ$, and
$l^\pm\mathord{\stackrel{{\scriptscriptstyle(}-{\scriptscriptstyle)}}{\nu}}\!_{l}\PW^\mp$
vertex functions can be found in \citere{Hollik:1988ii}.

All pentagon and box diagrams as well as the $\PH\gamma\gamma$ vertex
function are UV finite; also the $\PH\nu_l\bar{\nu}_l$ and $\PH
l^+l^-$ vertex functions are UV finite since we neglect the masses of
the light fermions everywhere apart from the mass-singular logarithms.
For the other vertex functions, $\PH\PW\PW$, $\PH\PZ\PZ$, $\PH\PZ\ga$,
$\nu_l \bar{\nu}_l\PZ$, $l^-l^+\PZ$,
$l^\pm\mathord{\stackrel{{\scriptscriptstyle(}-{\scriptscriptstyle)}}{\nu}}\!_{l}\PW^\mp$,
and for the relevant self-energies the corresponding counterterm
diagrams have to be included.

\subsection{Calculation of the one-loop corrections}

\subsubsection{Algebraic reduction of diagrams and standard matrix elements}
\label{se:alg1loop}

The algebraic part of the two calculations has been carried out in the
same way as in the one-loop calculation of
$\Pep\Pem\to\nu\bar\nu\PH$ described in \citere{Denner:2003yg}.
This means that 
we separate the fermion spinor chains from the rest of the
amplitude by defining standard matrix elements (SME).  
To introduce a compact notation for the SME, the tensors 
\beqar
\Gamma^{ab,\si}_{\{\al,\al\be\ga\}} &=&
\bar u_{f_a}(k_a)\left\{\ga_\al,\ga_\al\ga_\be\ga_\ga\right\}
\omega_\si v_{\bar f_b}(k_b),
\nn\\
\Gamma^{cd,\tau}_{\{\al,\al\be\ga\}} &=&
\bar u_{f_c}(k_c)\left\{\ga_\al,\ga_\al\ga_\be\ga_\ga\right\}
\omega_\tau v_{\bar f_d}(k_d)
\eeqar
are defined with obvious notations for the Dirac spinors $\bar
u_{f_a}(k_a)$, etc., and $\omega_\pm=(1\pm\gamma_5)/2$ denote the
right- and left-handed chirality projectors.  Here and in the
following, each entry in the set within curly brackets refers to a
single object, i.e.\ from the first line in the equation above we have
$\Gamma^{ab,\si}_{\al} = \bar u_{f_a}(k_a)\ga_\al \omega_\si v_{\bar
  f_b}(k_b)$ and $\Gamma^{ab,\si}_{\al\be\ga} = \bar
u_{f_a}(k_a)\ga_\al\ga_\be\ga_\ga \omega_\si v_{\bar f_b}(k_b)$,
 etc. Furthermore, symbols like $\Gamma_p$ are used as
shorthand for the contraction $\Gamma_\mu\, p^\mu$. 
We define the 52 SME
\newcommand{\Msme}{\hat\M}
\beq
\def\arraystretch{1.5}
\begin{array}[b]{rclcrcl}
\Msme^{abcd,\si\tau}_{\{1,2\}} &=&
\Ga^{ab,\si}_{\al} \; \Ga^{cd,\tau,\{\al,\al{k}_a{k}_b\}},
& \qquad &
\Msme^{abcd,\si\tau}_{\{3,4\}} &=&
\Ga^{ab,\si}_{\al k_c k_d} \; \Ga^{cd,\tau,\{\al,\al{k}_a{k}_b\}},
\\
\Msme^{abcd,\si\tau}_{\{5,6\}} &=&
\Ga^{ab,\si}_{k_c} \; \Ga^{cd,\tau,\{{k}_a,{k}_b\}},
& \qquad &
\Msme^{abcd,\si\tau}_{\{7,8\}} &=&
\Ga^{ab,\si}_{k_d} \; \Ga^{cd,\tau,\{{k}_a,{k}_b\}},
\\
\Msme^{abcd,\si\tau}_{\{9,10\}} &=&
\Ga^{ab,\si}_{\al\be k_c} \; \Ga^{cd,\tau,\{\al\be {k}_a,\al\be {k}_b\}},
& \qquad &
\Msme^{abcd,\si\tau}_{\{11,12\}} &=&
\Ga^{ab,\si}_{\al\be k_d} \; \Ga^{cd,\tau,\{\al\be {k}_a,\al\be {k}_b\}},
\\
\Msme^{abcd,\si\tau}_{13} &=&
\Ga^{ab,\si}_{\al\be\ga} \; \Ga^{cd,\tau,\al\be\ga}.
&&&&
\end{array}
\eeq
The SME are evaluated within the WvdW spinor technique, similar to the
lowest-order amplitudes described in the previous section.  The
tree-level and one-loop amplitudes $\M^{abcd,\si\tau}_0$ and
$\M^{abcd,\si\tau}_1$, respectively, for the generic four-fermion
final state $f_a\bar f_b f_c\bar f_d$ can be expanded in terms of
linear combinations of SME,
\beq
\M^{abcd,\si\tau}_n = 
\sum_{i=1}^{13} F^{abcd,\si\tau}_{n,i} \Msme^{abcd,\si\tau}_i,
\qquad n=0,1,
\eeq
with Lorentz-invariant functions $F^{abcd,\si\tau}_{n,i}$.
In this notation the lowest-order amplitudes \refeq{eq:MHffff-spec}
read
\beqar
\M_0^{\ZZ,\si_a\si_b\si_c\si_d}(k_a,k_b,k_c,k_d) &=&
\frac{e^3g^{\si_a}_{\PZ f_af_b}g^{\si_c}_{\PZ f_cf_d}\cmw}{\cw^2\sw} \,
\de_{\si_a,-\si_b}\de_{\si_c,-\si_d}
\nn\\ && {} \times
\frac{1}{[(k_a+k_b)^2-\cmz^2] [(k_c+k_d)^2-\cmz^2]}\,
\Msme^{abcd,\si_a\si_c}_1,
\nl
\M_0^{\WW,\si_a\si_b\si_c\si_d}(k_a,k_b,k_c,k_d) &=&
 \frac{e^3\cmw}{2\sw^3} \,
\de_{\si_a,-}\de_{\si_b,+}\de_{\si_c,-}\de_{\si_d,+}
\nn\\ && {} \times
\frac{1}{[(k_a+k_b)^2-\cmw^2] [(k_c+k_d)^2-\cmw^2]}\,
\Msme^{abcd,--}_1.
\hspace{2em}
\eeqar
For the one-loop amplitudes in general all invariant functions receive
contributions, in particular, they contain the loop integrals.  The
one-loop amplitudes for the various final states are constructed from
the amplitudes for $\PH\to f\bar f F\bar F$ and $\PH\to f\bar f' F\bar
F'$ as described in \refeq{eq:zz} to \refeq{eq:mixed} for the lowest
order.  The one-loop correction to the partial decay widths, finally,
reads
\beq
\int\rd\Ga_{\mathrm{virt}} = \frac{1}{2\MH} \int \rd\Phi_0 \,
 \sum_{\si_1,\si_2,\si_3,\si_4=\pm\frac{1}{2}} 
2\Re\left\{ \M^{\si_1,\si_2,\si_3,\si_4}_1
(\M^{\si_1,\si_2,\si_3,\si_4}_0)^* \right\}.
\label{eq:vcs}
\eeq

The actual calculation of the one-loop diagrams has been carried out
in the 't~Hooft--Feynman gauge. The Feynman graphs are evaluated in
two completely independent ways, leading to two independent computer
codes. The results of the two codes are in good numerical agreement
(i.e.\ within more than 10 digits for non-exceptional phase-space
points).

In the first calculation, the Feynman graphs are generated with {\sl
  Feyn\-Arts} version 1.0 \cite{Kublbeck:1990xc}.  With the help of
{\sl Mathematica} routines the amplitudes are expressed in terms of
SME and coefficients of tensor integrals. The output is processed into
a {\sl Fortran} program for the numerical evaluation.  This
calculation of the virtual corrections has been repeated using the
background-field method \cite{Denner:1994xt}, where the individual
contributions from self-energy, vertex, and box corrections differ
from their counterparts in the conventional formalism. The total
one-loop corrections of the conventional and of the background-field
approach were found to be in perfect numerical agreement.

The second calculation has been made using {\sl FeynArts} version~3
\cite{Hahn:2000kx} for the generation and {\sl FormCalc}
\cite{Hahn:1998yk} for the evaluation of the amplitudes.  The
analytical results of {\sl FormCalc} in terms of Weyl-spinor chains
and their coefficients have been translated to {\sl C++} code for the
numerical evaluation.

\subsubsection{Gauge-boson resonances and complex-mass scheme}
\label{se:cms}

The description of resonances in (standard) perturbation theory
requires a Dyson summation of self-energy insertions in the resonant
propagator in order to introduce the imaginary part provided by the
finite decay width into the propagator denominator. This procedure in
general violates gauge invariance, \ie destroys Slavnov--Taylor or
Ward identities and disturbs the cancellation of gauge-parameter
dependences, because different perturbative orders are mixed (see, for
instance, \citere{Grunewald:2000ju} and references therein).

In both of our two calculations we employ the so-called ``complex-mass
scheme'', which was introduced in \citere{Denner:1999gp} for
lowest-order calculations and generalized to the one-loop level in
\citere{Denner:2005es}.  In this approach the W- and Z-boson masses
are consistently considered as complex quantities, defined as the
locations of the propagator poles in the complex plane.  To this end,
bare real masses are split into complex renormalized masses and
complex counterterms.  Since the bare Lagrangian is not changed,
double counting does not occur.  Perturbative calculations can be
performed as usual, only parameters and counterterms, in particular
the electroweak mixing angle defined from the ratio of the W- and
Z-boson masses, become complex. Since we only perform an analytic
continuation of the parameters, all relations that follow from gauge
invariance, such as Ward identities, remain valid. As a consequence
the amplitudes are gauge independent, and unitarity cancellations are
respected.  Moreover, the on-shell renormalization scheme can
straightforwardly be transferred to the complex-mass scheme
\cite{Denner:2005es}.

The use of complex gauge-boson masses necessitates the consistent use
of these complex masses also in loop integrals. The scalar master
integrals are evaluated for complex masses using the methods and
results of Refs.~\cite{'tHooft:1978xw,Beenakker:1988jr,Denner:1991qq}.

We also treat the width of the top quark in the complex-mass scheme,
\ie we introduce a complex top-quark mass $\cmt$ via
$\cmt^2=\Mt^2-\ri\Mt\Gt$.

\subsubsection{Numerically stable evaluation of one-loop integrals}

The one-loop calculation of the decay $\PH\to4f$ requires the
evaluation of 5-point one-loop tensor integrals.  We calculate the
5-point integrals by directly reducing them to five 4-point functions,
as described in \citeres{Denner:2002ii,Denner:2005nn}. Note that this
reduction does not involve inverse Gram determinants composed of
external momenta, which naturally occur in the Passarino--Veltman
reduction \cite{Passarino:1979jh} of tensor to scalar integrals. The
latter procedure leads to serious numerical problems when the Gram
determinants become small.

Tensor 4-point and 3-point integrals are reduced to scalar integrals
with the Passarino--Veltman algorithm \cite{Passarino:1979jh} as long
as no small Gram determinant appears in the reduction. If small Gram
determinants occur, two alternative schemes are applied
\cite{Denner:2005nn}.  In one method, we evaluate a specific tensor
coefficient, the integrand of which is logarithmic in Feynman
parametrization, by numerical integration. Then the remaining
coefficients as well as the standard scalar integral are algebraically
derived from this coefficient.  This method is used in the first loop
calculation described in \refse{se:alg1loop}.  The second, alternative
method, which is used in the second loop calculation described in
\refse{se:alg1loop}, makes use of expansions of the tensor
coefficients about the limit of vanishing Gram determinants and
possibly other kinematical determinants. In this way, all tensor
coefficients can be expressed in terms of the standard scalar
functions.

The whole procedure for the evaluation of the scalar and tensor
one-loop integrals has been taken over from the one-loop calculation
of $\Pep\Pem\to4\,$fermions \cite{Denner:2005es}.

\subsubsection{Input-parameter scheme}
\label{se:inputscheme}

We use the ``\GF{} scheme'', where a large universal part of the $\Oa$
corrections is absorbed into the lowest-order prediction.  In this
scheme the electromagnetic coupling constant $\al=e^2/(4\pi)$ is
derived from the Fermi constant $\GF$, the muon decay constant,
according to
\beq\label{eq:al-GF}
\alpha_{\GF}=\frac{\sqrt{2}\GF\MW^2}{\pi}\left(1-\frac{\MW^2}{\MZ^2}\right). 
\eeq
This procedure takes into account the running of the electromagnetic
coupling constant $\al(Q^2)$ from $Q^2=0$ to the electroweak scale and
also accounts for universal corrections related to the $\rho$
parameter in the coupling of the W~boson to fermions.  

In order to avoid double-counting, the corrections absorbed in the
lowest-order prediction by the use of $\alpha_{\GF}$ have to be
subtracted from the explicit $\Oa$ corrections. To this end, we
subtract the weak corrections to muon decay $\Delta r$
\cite{Denner:1993kt,Sirlin:1980nh} from the corrections in the
$\alpha(0)$ (on-shell) scheme.  This can be done by redefining the
charge renormalization constant as
\beq
\de Z_e\Big|_{\GF} = \de Z_e\Big|_{\al(0)} - \frac{1}{2} (\Delta
r)_{\mbox{\scriptsize 1-loop}},
\eeq
where $(\Delta r)_{\mbox{\scriptsize 1-loop}}$ is the one-loop
expression for $\Delta r$ evaluated in the complex-mass scheme.

\subsection{Leading two-loop corrections}
\label{se:gf2mh4}

Since corrections due to the self-interaction of the Higgs boson
become important for large Higgs masses, we have included the dominant
two-loop corrections to the decay $\PH\to\PV\PV$ in the
large-Higgs-mass limit which were calculated in
\citeres{Ghinculov:1995bz,Frink:1996sv}.  They are of order
$\Ord(\GF^2\MH^4)$ and read
\beq\label{eq:GF2MH4}
\int\rd\Gamma_{\GF^2\MH^4} = 
62.0308(86) \left(\frac{\GF\MH^2}{16\pi^2\sqrt{2}}\right)^2
\int\rd\Gamma_0,
\eeq
where the numerical prefactor has been taken from
\citere{Frink:1996sv}.  The error of this factor is far beyond other
uncertainties and, thus, ignored in the numerics.

We do not include any higher-order corrections proportional to a power
of $\GF\Mt^2$ since we already see at the one-loop level that the
heavy-top limit does not provide a sound approximation of the
corrections from closed fermion loops.  In particular, for Higgs
masses near and above the $\Pt\bar\Pt$ threshold, $\MH\gsim2\Mt$ the
large-$\Mt$ limit is not appropriate.

\section{Real photon corrections}
\label{se:real}

\subsection{Matrix element for $\PH\to4f\gamma$}
\label{se:calcrcs}

The real photon corrections are induced by the process
\beq
\PH(p)
 \;\longrightarrow\;
f_1(k_1,\si_1) + \bar f_2(k_2,\si_2) + f_3(k_3,\si_3) + \bar
f_4(k_4,\si_4) + \gamma(k,\lambda),  
\label{eq:h4fa}
\eeq
where the momenta and helicities of the external particles are
indicated in parentheses.

As for the lowest-order process, we consistently neglect fermion
masses. However, we restore the mass-singular logarithms appearing in
collinear photon emission as described in \refse{se:softcoll}.

The matrix elements for the radiative process can be constructed in
the same way as for the lowest-order process \refeq{process-H4f} 
from the set of generic diagrams that is obtained from
\reffi{fi:H4f-born-diag} by adding a photon line in all possible ways to
the charged particles 
(including possible new graphs involving would-be Goldstone-boson 
exchange).

We have evaluated the generic helicity matrix elements
$\M^{\si_a\si_b\si_c\si_d\la}_\ga(k_a,k_b,k_c,k_d,k)$ of this process
again using the WvdW spinor technique in the formulation of
\citere{Dittmaier:1998nn}.  The amplitudes generically read
\beqar\label{eq:MEH4fa}
\lefteqn{\M^{VV,\si_a\si_b\si_c\si_d\la}_{\ga}(Q_a,Q_b,Q_c,Q_d,k_a,k_b,k_c,k_d,k) =} 
\qquad\\
&&
2\sqrt{2}e^4\,
g^{\si_a}_{Vf_af_b}g^{\si_c}_{Vf_cf_d}g_{\PH VV} \,
 \de_{\si_a,-\si_b}\de_{\si_c,-\si_d}\,
A_{\si_a\si_c\la}^{VV}(Q_a,Q_b,Q_c,Q_d,k_a,k_b,k_c,k_d,k),\nn
\eeqar
or more specifically for the case of $\PZ$-mediated and $\PW$-mediated
decays
\beqar
\lefteqn{\M^{\ZZ,\si_a\si_b\si_c\si_d\la}_{\ga}(Q_a,Q_b,Q_c,Q_d,k_a,k_b,k_c,k_d,k) =} 
\qquad\nl
&&
\frac{2\sqrt{2}e^4g^{\si_a}_{\PZ f_af_b}g^{\si_c}_{\PZ f_cf_d}\cmw}{\cw^2\sw} \,
\de_{\si_a,-\si_b}\de_{\si_c,-\si_d}\,
A_{\si_a\si_c\la}^{\ZZ}(Q_a,Q_b,Q_c,Q_d,k_a,k_b,k_c,k_d,k),
\nn\\[.5em]
\lefteqn{\M^{\WW,\si_a\si_b\si_c\si_d\la}_{\ga} (Q_a,Q_b,Q_c,Q_d,k_a,k_b,k_c,k_d,k) =} 
\qquad\nl
&&
 \frac{\sqrt{2}e^4\cmw}{\sw^3} \,
\de_{\si_a,-}\de_{\si_b,+}\de_{\si_c,-}\de_{\si_d,+}\,
A_{--\la}^{\WW}(Q_a,Q_b,Q_c,Q_d,k_a,k_b,k_c,k_d,k).
\eeqar
The auxiliary functions are given by
\beqar\label{eq:MEH4faaux}
\lefteqn{A_{---}^{VV}(Q_a,Q_b,Q_c,Q_d,k_a,k_b,k_c,k_d,k) =}\qquad\nl&&
\langle k_b k_d\rangle^*\Biggl[
\frac{\langle k_a k_b \rangle^* \langle k_a k_c\rangle
      +\langle k k_b   \rangle^* \langle k k_c  \rangle}
{[(k_a+k_b+k)^2-\cmvs][(k_c+k_d)^2-\cmvs]}\nl*
&&\qquad \times\left(\frac{Q_a}{\langle k k_a   \rangle^* \langle k k_b
    \rangle^*}
+\frac{Q_a-Q_b}{(k_a+k_b)^2-\cmvs}
\frac{\langle k k_a \rangle}{\langle k k_b \rangle^*}
\right)
\nl&&{}-
\frac{\langle k_c k_d \rangle^* \langle k_c k_a\rangle
      +\langle k k_d   \rangle^* \langle k k_a  \rangle}
{[(k_a+k_b)^2-\cmvs][(k_c+k_d+k)^2-\cmvs]}\nl
&&\qquad \times\left(\frac{Q_c}{\langle k k_c   \rangle^* \langle k k_d
    \rangle^*}
+\frac{Q_c-Q_d}{(k_c+k_d)^2-\cmvs}
\frac{\langle k k_c \rangle}{\langle k k_d \rangle^*}
\right)
\nl&&{}+
\frac{Q_a-Q_b}
{[(k_a+k_b)^2-\cmvs][(k_c+k_d)^2-\cmvs]}
\frac{\langle k_b k_d \rangle^* \langle k_a k_c \rangle}
     {\langle k k_b \rangle^* \langle k k_d \rangle^*}
\biggr],
\hspace{6em}
\nl
\lefteqn{A_{+--}^{VV}(Q_a,Q_b,Q_c,Q_d,k_a,k_b,k_c,k_d,k) = 
A_{---}^{VV}(-Q_b,-Q_a,Q_c,Q_d,k_b,k_a,k_c,k_d,k),}\qquad\nl
\lefteqn{A_{-+-}^{VV}(Q_a,Q_b,Q_c,Q_d,k_a,k_b,k_c,k_d,k) = 
A_{---}^{VV}(Q_a,Q_b,-Q_d,-Q_c,k_a,k_b,k_d,k_c,k),}\qquad\nl
\lefteqn{A_{++-}^{VV}(Q_a,Q_b,Q_c,Q_d,k_a,k_b,k_c,k_d,k) = 
A_{---}^{VV}(-Q_b,-Q_a,-Q_d,-Q_c,k_b,k_a,k_d,k_c,k),}\qquad\nl
\lefteqn{A_{\si_a\si_c+}^{VV}(Q_a,Q_b,Q_c,Q_d,k_a,k_b,k_c,k_d,k) = }\qquad\nl&&
\Bigl(A_{-\si_a,-\si_c,-}^{VV}(Q_a,Q_b,Q_c,Q_d,k_a,k_b,k_c,k_d,k)\Bigr)^*
\Bigr|_{\cmv\to \cmv^*},
\eeqar
and obey the relations
\beqar
\lefteqn{A_{-\si_a,-\si_c,-\la}^{VV}(Q_a,Q_b,Q_c,Q_d,k_a,k_b,k_c,k_d,k) = }\qquad\nl&&
\Bigl(A_{\si_a\si_c\la}^{VV}(Q_a,Q_b,Q_c,Q_d,k_a,k_b,k_c,k_d,k)\Bigr)^*
\Bigr|_{\cmv\to \cmv^*},\nl
\lefteqn{A_{-\si_a,\si_c,\la}^{VV}(Q_a,Q_b,Q_c,Q_d,k_a,k_b,k_c,k_d,k) = }\qquad\nl&&
A_{\si_a\si_c\la}^{VV}(-Q_b,-Q_a,Q_c,Q_d,k_b,k_a,k_c,k_d,k), \nl
\lefteqn{A_{\si_a,-\si_c,\la}^{VV}(Q_a,Q_b,Q_c,Q_d,k_a,k_b,k_c,k_d,k) = }\qquad\nl&&
A_{\si_a\si_c\la}^{VV}(Q_a,Q_b,-Q_d,-Q_c,k_a,k_b,k_d,k_c,k),\nl
\lefteqn{A_{\si_a,\si_c,-\la}^{VV}(Q_a,Q_b,Q_c,Q_d,k_a,k_b,k_c,k_d,k) =}\qquad\nl&&
-\Bigl(A_{\si_c\si_a\la}^{VV}(Q_d,Q_c,Q_b,Q_a,k_d,k_c,k_b,k_a,k)\Bigr)^*\Bigr|_{\cmv\to \cmv^*},
\nl
\lefteqn{A_{\si_a\si_c\la}^{VV}(Q_a,Q_b,Q_c,Q_d,k_a,k_b,k_c,k_d,k) = }\qquad\nl&&
A_{\si_c\si_a\la}^{VV}(Q_c,Q_d,Q_a,Q_b,k_c,k_d,k_a,k_b,k)
.
\eeqar
The relations between the $A^{\dots}_{\dots}$ functions that differ in
all helicities result from a P transformation. Those, where only one
fermion helicity is reversed are related to C symmetry.  
The last but one is due to CP symmetry, and the last one results from a
symmetry under the exchange of the two fermion pairs.
The charges of the fermions are related by
\beq
Q_a-Q_b+Q_c-Q_d=0.
\eeq

For the $\PZ$-mediated decays, where $Q_a=Q_b$ and $Q_c=Q_d$, the
auxiliary function \refeq{eq:MEH4faaux} simplifies to
\beqar
\lefteqn{A_{---}^{\ZZ}(Q_a,Q_a,Q_c,Q_c,k_a,k_b,k_c,k_d,k) = }\qquad\nl*
\qquad&& \langle k_b k_d\rangle^*\Biggl[
\frac{\langle k_a k_b \rangle^* \langle k_a k_c\rangle
      +\langle k k_b   \rangle^* \langle k k_c  \rangle}
{[(k_a+k_b+k)^2-\cmvs][(k_c+k_d)^2-\cmvs]}
\frac{Q_a}{\langle k k_a   \rangle^* \langle k k_b
    \rangle^*} 
\nl\qquad&&{}-
\frac{\langle k_c k_d \rangle^* \langle k_c k_a\rangle
      +\langle k k_d   \rangle^* \langle k k_a  \rangle}
{[(k_a+k_b)^2-\cmvs][(k_c+k_d+k)^2-\cmvs]}
\frac{Q_c}{\langle k k_c   \rangle^* \langle k k_d
    \rangle^*}
\biggr].
\eeqar

From the generic matrix element
$\M_\ga^{VV,\si_a\si_b\si_c\si_d\la}(k_a,k_b,k_c,k_d,k)$ the matrix
elements for the specific processes can be constructed in complete
analogy to the process without photon as in
\refeq{eq:zz}--\refeq{eq:mixed}.

The squares of the matrix elements \refeq{eq:MEH4fa} have been
successfully checked against the result obtained with the package 
{\sc Madgraph} \cite{Stelzer:1994ta} numerically.

The contribution $\Ga_\gamma$ of the radiative decay to the total
decay width is given by
\beq
\int\rd\Ga_\gamma = \frac{1}{2\MH} \int \rd\Phi_\gamma \,
 \sum_{\si_1,\si_2,\si_3,\si_4=\pm\frac{1}{2}}
\, \sum_{\lambda=\pm 1} \,
|\M^{\si_1\si_2\si_3\si_4\la}_\gamma|^2,
\label{eq:hbcsga}
\eeq
where the phase-space integral is defined by
\beq
\int \rd\Phi_\gamma =
\int\frac{\rd^3 {\bf k}}{(2\pi)^3 2k^0} \,
\left( \prod_{i=1}^4 \int\frac{\rd^3 {\bf k}_i}{(2\pi)^3 2k_i^0} \right)\,
(2\pi)^4 \delta^{(4)}\Biggl(p-k-\sum_{j=1}^4 k_j\Biggr).
\label{eq:dPSg}
\eeq
Without introducing soft and collinear regulators this phase-space integral 
diverges in the soft ($k^0\to0$) and collinear ($kk_i\to0$) regions.
The applied solutions are outlined in the next section.

\subsection{Treatment of soft and collinear divergences}
\label{se:softcoll}

In the combination of virtual and real photon corrections, the
fermion-mass effects have to be restored in the phase-space
regions of collinear photon radiation off charged fermions, and the IR
regularization for soft-photon emission has to be implemented. To this
end, we employ the dipole subtraction formalism for photon radiation
\cite{Dittmaier:2000mb} as well as the more conventional phase-space
slicing approach.

The actual calculation exactly follows the one described in
\citere{Bredenstein:2005zk}, where electroweak corrections to the
related process $\ga\ga\to\PW\PW\to4f$ have been calculated.
The structure of soft and collinear singularities of this process 
is exactly the same as in the decay $\PH\to4f$ considered in this work,
because both processes involve the same pattern of charged particles
in the initial and final states. Consequently, apart from obvious
substitutions for the 
flux factors all formulas given in Section~4 of \citere{Bredenstein:2005zk}
for cross sections literally carry over to our decay widths.

In the matching of real and virtual corrections the issue of
collinear safety of observables is crucial.
We speak of collinear-safe observables
if a nearly collinear system of a charged fermion and a photon is treated
inclusively, i.e.\ if phase-space selection cuts
(or histogram bins of distributions)
depend only on the sum $k_i+k$ of the nearly collinear
fermion and photon momenta. In this case the energy fraction
\beq
z_i=\frac{k^0_i}{k^0_i+k^0}
\label{eq:zi}
\eeq
of a charged fermion $f_i$ after emitting a photon in a sufficiently
small cone around its direction of flight is fully integrated over,
because it is not constrained by any phase-space cut (or histogram bin
selection in distributions).  Thus, the Kinoshita--Lee--Nauenberg
(KLN) theorem \cite{Kinoshita:1962ur} guarantees that all
singularities connected with FSR cancel between the virtual and real
corrections, even though they are defined on different phase spaces.
The full phase-space integration in $\PH\to4f(+\ga)$, which leads to
partial decay widths, is trivially collinear safe.  More generally, a
sufficient inclusiveness is achieved by the photon recombination
described in \refse{se:input}, which treats outgoing charged fermions
and photons as one quasi-particle if they are very close in angle.
The original version of the dipole subtraction formalism for photon
radiation \cite{Dittmaier:2000mb} deals with collinear-safe situations
only.  The generalization to the non-collinear-safe FSR is described
in \citere{Bredenstein:2005zk}.

\begin{figure}
\setlength{\unitlength}{1cm}
\centerline{
\begin{picture}(7.7,8)
\put(-1.7,-14.5){\includegraphics{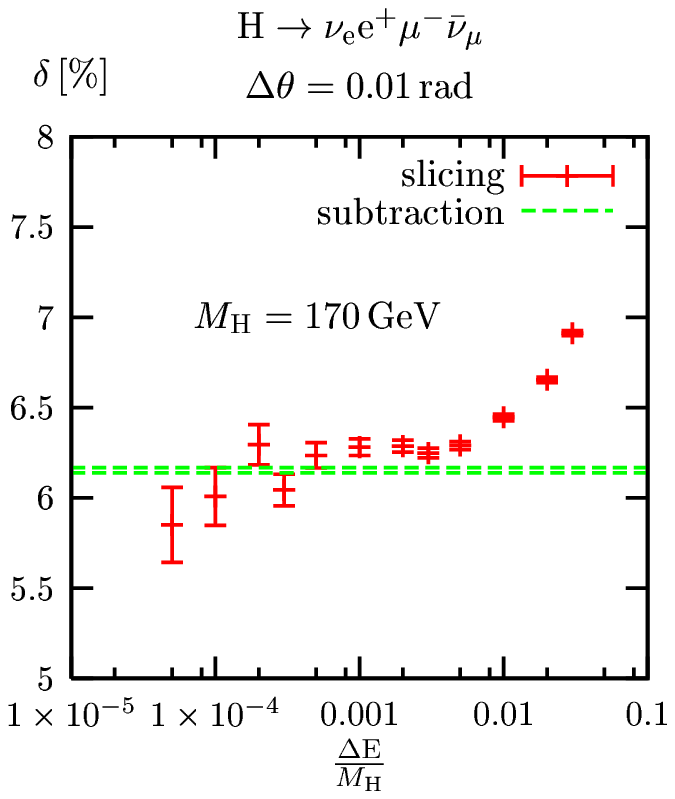}}
\end{picture}
\begin{picture}(7.5,8)
\put(-1.7,-14.5){\includegraphics{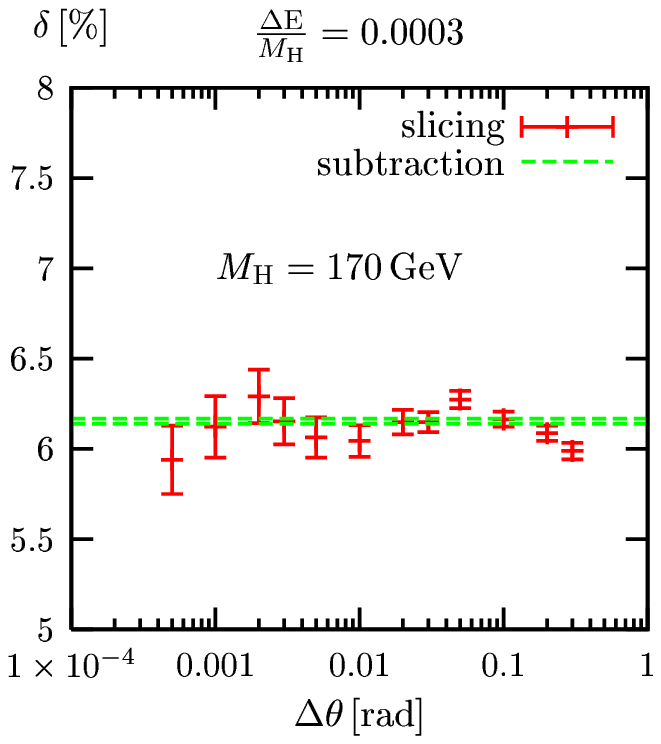}}
\end{picture} }
\caption{Dependence of the relative corrections $\de$ to the partial
  decay width on the energy cutoff $\Delta E$ (l.h.s.)\ and on the
  angular cutoff $\Delta\theta$ (r.h.s.)\  in the slicing approach for
  the decay $\PH\to\Pne\Pep\mu^-\bar\nu_\mu$ with $\MH=170\GeV$. For
  comparison the corresponding result obtained with the dipole
  subtraction method is shown as a $1\si$ band in the plots.}
\label{fig:sliWW}
\vspace*{2em}
\setlength{\unitlength}{1cm}
\centerline{
\begin{picture}(7.7,8)
\put(-1.7,-14.5){\includegraphics{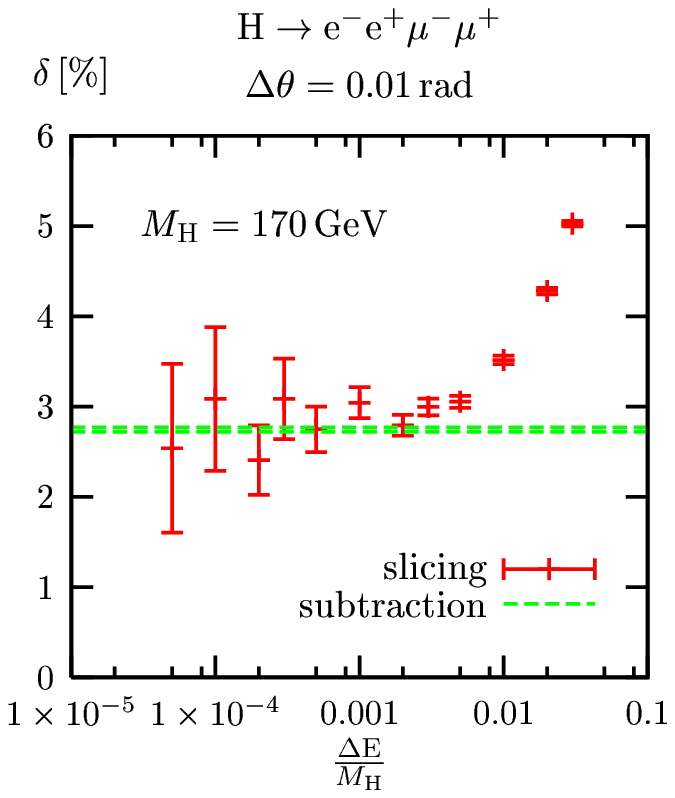}}
\end{picture}
\begin{picture}(7.5,8)
\put(-1.7,-14.5){\includegraphics{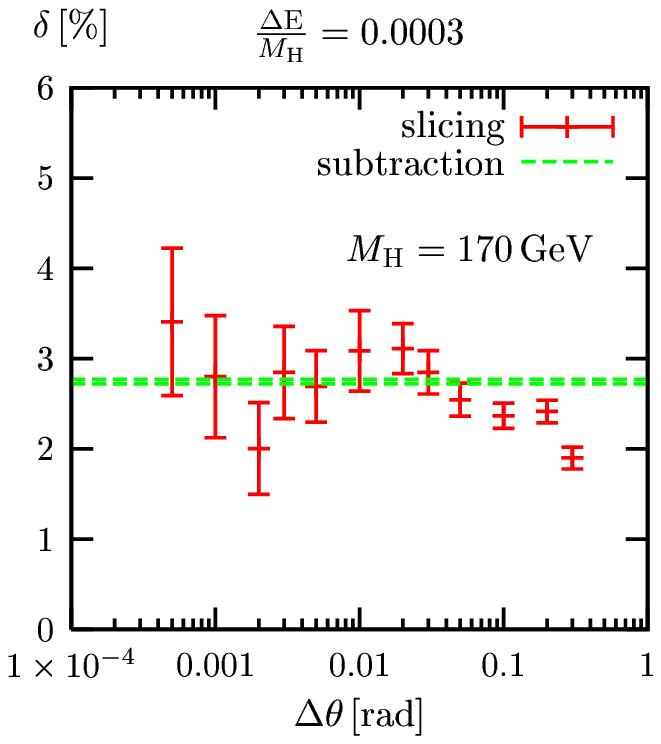}}
\end{picture} }
\caption{Same as in \reffi{fig:sliWW} but for the decay
$\PH\to\Pem\Pep\mu^-\mu^+$.}
\label{fig:sliZZ}
\end{figure}

Figures~\ref{fig:sliWW} and \ref{fig:sliZZ} illustrate the agreement
between the subtraction and slicing methods for the partial decay
widths of the two decay channels $\PH\to\Pne\Pep\mu^-\bar\nu_\mu$ and
$\PH\to\Pem\Pep\mu^-\mu^+$.
These results were obtained with $5\cdot10^7$ events.
In the slicing approach, the phase-space regions of soft or collinear
photons are defined by the auxiliary cutoff parameters 
$\De E\ll\GW$ and $\De\theta\ll1$.  
The region $k^0<\De E$ is treated in soft-photon
approximation, the regions $\theta_{\ga f}<\De\theta$, $k^0>\De E$
($\theta_{\ga f}$ is the emission angle from any fermion $f$) are
evaluated using collinear factorization. In the remaining regular
phase space no regulators (photon and fermion masses) are used.
Therefore, the slicing result is correct up to terms of ${\cal O}(\De
E)$ and ${\cal O}(\De\theta)$.  For decreasing auxiliary parameters
$\De E$ and $\De\theta$, the slicing result reaches a plateau, as it
should be, until the increasing statistical errors become large and
eventually underestimated.  In the plateau region the slicing and
subtraction
results are compatible within statistical errors, but the subtraction
result shows smaller integration errors although the same number of
events is used.

\subsection{Higher-order final-state radiation}
\label{se:fsr}
\newcommand{\isrscale}{M}

Photons that are emitted collinear off a charged fermion give 
rise to corrections that are enhanced by large logarithms of the form
$\al\log{m_f^2/\isrscale^2}$, where $m_f$ is a fermion mass and $\isrscale$ is some
typical energy scale. If the photons are treated fully inclusively, as
it is the case if the photons are recombined with the corresponding
fermion, these logarithms cancel due to the KLN theorem
\cite{Kinoshita:1962ur}.  If, however, distributions like in the
invariant mass of two fermions, as discussed in \refse{se:numerics},
are to be considered without recombining collinear photons, then these
logarithms do not cancel and yield large effects.  Thus, corrections
of this origin should be taken into account beyond $\Oa$.  This can be
achieved in the structure-function approach \cite{sf} which is based
on the mass-factorization theorem. According to this theorem the decay
width including the leading-logarithmic FSR terms can be written as
\beq
\int\rd\Ga_{\LLFSR} = \prod_{i=1 \atop Q_i\ne 0}^4\left[\int_0^1\rd z_i\, 
\Ga_{ii}^{\LL}(z_i,\isrscale^2)\right]\int\rd\Ga_0\,
\Theta_{\mathrm{cut}}(\{z_jk_j\}).
\label{eq:FSR}
\eeq
The function $\Theta_{\mathrm{cut}}(\{z_jk_j\})$ 
generically denotes all histogram routines or phase-space cuts. 
It depends on the fermion momenta $z_jk_j$ 
which, in the case of 
charged fermions, may be reduced by the factor
$z_j$ due to collinear photon emission. 
For neutral fermions we have $z_j=1$. 
The structure functions including terms up to $\Oaaa$, 
improved by the exponentiation of the soft-photon parts, read
\cite{Beenakker:1996kt}
\newcommand{\rE}{\mathrm{E}}
\beqar
  \Gamma_{ii}^{\LL,\mathrm{exp}}(z,\isrscale^2) &=&    
    \frac{\exp\left(-\frac{1}{2}\beta_i\gamma_{\rE} +
        \frac{3}{8}\beta_i\right)}
{\Gamma\left(1+\frac{1}{2}\beta_i\right)}
    \frac{\beta_i}{2} (1-z)^{\frac{\beta_i}{2}-1} - \frac{\beta_i}{4}(1+z) 
\nn\\
&&  {} - \frac{\beta_i^2}{32} \biggl\{ \frac{1+3z^2}{1-z}\ln(z)
    + 4(1+z)\ln(1-z) + 5 + z \biggr\}
\nn\\
&&  {} - \frac{\beta_i^3}{384}\biggl\{
      (1+z)\left[6\Li(z)+12\ln^2(1-z)-3\pi^2\right] 
\nn\\
&& \quad\quad {}
+\frac{1}{1-z}\biggl[ \frac{3}{2}(1+8z+3z^2)\ln(z) 
+6(z+5)(1-z)\ln(1-z)
\nn\\
&& \quad\quad\quad {}
+12(1+z^2)\ln(z)\ln(1-z)-\frac{1}{2}(1+7z^2)\ln^2(z)
\nn\\
&& \quad\quad\quad  {}
+\frac{1}{4}(39-24z-15z^2)\biggr] \biggr\}
\label{eq:GammaFSR}
\eeqar
with $\gamma_{\rE}$ and $\Gamma(y)$ denoting Euler's constant and
the Gamma function, respectively. The mass-singular logarithm
\beq
\beta_i = \frac{2Q_i^2 \alpha(0)}{\pi} 
\left[\ln\biggl(\frac{\isrscale^2}{m_i^2}\biggr)-1\right]
\eeq
involves a scale $\isrscale$, which is not fixed in leading
logarithmic order and should be set to a scale typical for the process
under consideration. We use $\isrscale^2=\MH^2$ in our evaluations.
The appropriate coupling in the leading-logarithmic terms is $\al(0)$,
since these originate from real or virtual soft or collinear photons.
As the function $(1-z)^{\frac{\beta_i}{2}-1}$ is difficult to
integrate numerically, an appropriate mapping has to be chosen.

In order to study the influence of the higher-order terms we
alternatively expand the exponential up to terms of $\Oaaa$,
yielding
\beqar
\Gamma_{ii}^{\LL}(z,\isrscale^2) &=&    
  \delta(1-z)
 +\Biggl[\, \frac{\beta_i}{4}\frac{1+z^2}{1-z}
\nn\\
&&  {} +\frac{\beta_i^2}{32} \biggl\{\frac{1+4z+z^2}{1-z}  -\frac{1+3z^2}{1-z}\ln(z)
    + 4\frac{1+z^2}{1-z}\ln(1-z)\biggr\}
\nn\\
&&  {} + \frac{\beta_i^3}{1536}\biggl\{
\frac{15+24z+15z^2}{1-z} 
- 4\pi^2\frac{1+3z^2} {1-z}
-6\frac{1+8z+3z^2}{1-z} \ln(z)
\nn\\
&& \quad\quad {}
+ 24\frac{1+4z+z^2}{1-z}\ln(1-z) 
      -24(1+z)\Li(z)
+ 48\frac{1+z^2}{1-z}\ln^2(1-z)
\nn\\
&& \quad\quad {}
-48\frac{1+z^2}{1-z}\ln(z)\ln(1-z)
+2\frac{1+7z^2}{1-z}\ln^2(z)
\biggr\}\,
\Biggr]_+,
\label{eq:FSRreexpand}
\eeqar
where the $[\dots]_+$ prescription is defined as usual,
\beq
\int_0^1\rd z\, \Big[f(z)\Big]_+ g(z) \equiv
\int_0^1\rd z\, f(z) \left[g(z)-g(1)\right],
\eeq

We convolute the lowest-order width according to \refeq{eq:FSR} and
add this to the result for the  $\Oa$-corrected width.  
In order to avoid double counting, 
we have to subtract
\beq
\int\rd\Ga_{\LLFSR,1} = \int\rd\Ga_0 + 
\int\rd\Ga_0\sum_{i=1\atop Q_i\ne0}^4
\left[\int_0^1\rd z_i\,\Ga_{ii}^{\LL,1}(z_i,\isrscale^2)
   \Theta_{\mathrm{cut}}(z_ik_i,\{k_{j\ne i}\})\right],
\label{eq:Oafsr}
\eeq
i.e.\ the leading logarithmic terms up to $\Oa$, from 
$\int\rd\Ga_{\LLFSR}$. They are defined by 
\beqar\label{eq:Oasf}
  \Ga_{ii}^{\LL,1}(z,\isrscale^2) &=&
  \frac{\beta_{i,\GF}}{4} \left(\frac{1+z^2}{1-z}\right)_+ .
\eeqar
Note that we have to subtract the $\Oa$ terms according to the scheme
that is applied for the virtual corrections.  Since we work in the
$\GF$ scheme, $\beta_{i,\GF}$ is proportional to $\al_{\GF}$, as
defined in \refse{se:numerics}.

\section{Final prediction and Monte Carlo integration}
\label{se:MC}

Summarizing all contributions to the differential decay width,
\refeq{eq:hbcs}, \refeq{eq:vcs}, \refeq{eq:GF2MH4},
\refeq{eq:hbcsga}, \refeq{eq:FSR}, and \refeq{eq:Oafsr},
we get the following prediction,
\beq
\int\rd \Ga =
\int\rd \Ga_0
+\int\rd \Ga_{\mathrm{virt}}
+\int\rd \Ga_{\GF^2\MH^4}
+\int\rd \Ga_\gamma
+\int\rd \Ga_{\LLFSR}
-\int\rd \Ga_{\LLFSR,1}.
\eeq

The phase-space integrations are performed using
the multi-channel Monte Carlo technique
\cite{Berends:1994pv} where the integrand is flattened by choosing
appropriate mappings of the pseudo-random numbers into the momenta of
the outgoing particles.  In more detail, the Monte Carlo part of {\sc
  Prophecy4f} builds upon the existing generators {\sc RacoonWW}
\cite{Denner:1999gp,Denner:2002cg} and {\sc Coffer$\gamma\gamma$}
\cite{Bredenstein:2005zk,Bredenstein:2004ef}.  The results obtained
with the multi-channel technique have been checked against a second
integration program based on the adaptive integration program {\sc
  Vegas} \cite{Lepage:1977sw}.
Although {\sc Vegas} already performs some adaption to the integrand
structure a further improvement of the numerical error has been
achieved by using mappings of the integration variables designed to
flatten the resonances.

The numerical results presented below have been obtained using $5\cdot
10^7$ events except for the plots showing the decay width as a
function of the Higgs mass which were calculated using $2\cdot 10^7$
events per point.  Since the virtual corrections (rendered finite by
adding the soft and collinear singularities from the real
corrections), and also their statistical error, are at least a factor
10 smaller than the lowest-order values for moderate Higgs masses, we
only evaluated the virtual corrections every 100th time, which
improves the run-time of the program but does not deteriorate the
overall statistical error. Soft and collinear singularities were
treated with the subtraction method in the results shown below.

\section{Improved-Born Approximation}
\label{se:IBA}

The electroweak corrections contain large contributions of universal
origin.  Besides final-state radiation, which is discussed in
\refse{se:fsr}, these consist, in particular, of the corrections
associated with the running of $\al$, corrections proportional to
$\Mt^2/\MW^2$, and corrections proportional to $\MH^2/\MW^2$. By
suitable parametrization of the lowest-order matrix elements, 
some of these
universal corrections can be incorporated in the lowest order, thus
reducing the remaining corrections. This does not only reduce the
$\Oa$ corrections but in general also the higher-order corrections.
Corrections associated with the running of $\al$, and corrections
related to the $\rho$ parameter in the W-boson--fermion coupling are
incorporated in the lowest-order prediction by using the $\GF$ scheme.

Some loop diagrams involving top quarks lead to corrections that are
enhanced by a large coupling factor $\GF\Mt^2$ in the limit of a large
top-quark mass $\Mt$. For the $\PH\PZ\PZ$ and $\PH\PW\PW$ vertices 
this type of corrections was considered even up to two-loop order
in \citere{Kniehl:1995at}. However, the pure heavy-top limit
$\Mt\to\infty$ is only applicable for $\MH\ll2\Mt$.
Since we are also interested in $\MH$ values near and above the 
$\Pt\bar\Pt$ threshold, we instead consider the more general
limit $\Mt,\MH\gg\MW,\MZ,m_{f\ne\Pt}$, 
i.e.\ we do not assume any hierarchy between
the Higgs and the top-quark masses.
We evaluate all closed fermion loops in this limit and keep only
contributions that are enhanced by $\Mt^2$ times any function
of the ratio $\MH/\Mt$. The numerical analysis shows that this
procedure yields a very good approximation for the sum of all
closed fermion loops in the $\GF$ scheme.
In order to approximate the remaining bosonic corrections,
the leading one- and two-loop corrections to the $\PH\PV\PV$ vertices
in the large-Higgs-mass limit, 
which are proportional to $\GF\MH^2$ and $\GF^2\MH^4$, respectively, 
are included in the IBA; these corrections are taken from
\citeres{Ghinculov:1995bz,Frink:1996sv}. 
Moreover,
for $\PH\to\PW\PW\to 4f$ we include the leading effect $\de_{\Coul}$ of
the Coulomb singularity as calculated in \citere{Fadin:1993kg}, 
which originates from soft-photon exchange
between the two slowly moving W~bosons near the WW~threshold.  
Finally, we take into account the QCD corrections to the gauge-boson
decays if quarks are involved in the final states.
The remaining corrections are expected to be widely independent of
Higgs mass $\MH$ and of the choice of the $4f$ final state.

Our IBA for the partial decay widths, which is constructed 
according to these lines, reads
\beqar
\int\rd\Gamma_{\IBA}^{\PH\to\PZ\PZ\to 4f}&=&
\frac{1}{2\MH}\int\rd\Phi_0\,\sum_{\si_1,\si_2,\si_3,\si_4=\pm}
|\M^{\ZZ,\si_1\si_2\si_3\si_4}_0|^2
\nl&&\qquad{}
\times \Re\left\{1+
\frac{\GF\cmt^2}{8\sqrt{2}\pi^2}
\Biggl[1-\frac{6\cw}{\sw}
\left(\frac{Q_{f_1}}{g^{\si_1}_{\PZ f_1 f_1}}
+\frac{Q_{f_3}}{g^{\si_3}_{\PZ f_3 f_3}}\right)
+\tau_{\PH\PZ\PZ}\left(\frac{\MH^2}{\cmt^2}\right)
\right]
\nl&&\qquad\qquad{}
+\frac{\GF\MH^2}{8\sqrt{2}\pi^2}
\left(\frac{5\pi^2}{6}-3\sqrt{3}\pi+\frac{19}{2} \right)
+62.0308(86) \left(\frac{\GF\MH^2}{16\pi^2\sqrt{2}}\right)^2
\nl&&\qquad\qquad{}
+\de_{\PZ\to f_1\bar f_2}^{\QCD}
+\de_{\PZ\to f_3\bar f_4}^{\QCD}
+c_{\PH\PZ\PZ}
\Biggr\},
\nn\\[1em]
\int\rd\Gamma_{\IBA}^{\PH\to\PW\PW\to 4f}&=&
\frac{1}{2\MH}\int\rd\Phi_0\,
|\M^{\WW,-+-+}_0|^2
\nl&&\qquad{}
\times \Re\Biggl\{1+
\frac{\GF\cmt^2}{8\pi^2\sqrt{2}}
\left[-5 +\tau_{\PH\PW\PW}\left(\frac{\MH^2}{\cmt^2}\right)
\right]
\nl&&\qquad\qquad{}
+\frac{\GF\MH^2}{8\sqrt{2}\pi^2}
\left(\frac{5\pi^2}{6}-3\sqrt{3}\pi+\frac{19}{2} \right)
+62.0308(86) \left(\frac{\GF\MH^2}{16\pi^2\sqrt{2}}\right)^2
\nl&&\qquad\qquad{}
+g(\bar\beta)
\de_{\Coul}\left(\MH^2,(k_1+k_2)^2,(k_3+k_4)^2\right) \, 
\nl&&\qquad\qquad{}
+\de_{\PW\to f_1\bar f_2}^{\QCD}
+\de_{\PW\to f_3\bar f_4}^{\QCD}
+c_{\PH\PW\PW}
\Biggr\},
\label{eq:IBA}
\eeqar
where the terms proportional to a charge factor $Q_f$ are absent if
$f$ is a neutrino.  The phase-space integral was defined in
\refeq{eq:dPS}.  The auxiliary functions $\tau_{\PH\PV\PV}$, which
appear in \refeq{eq:IBA}, are given by
\beqar
\tau_{\PH\PZ\PZ}\left(\frac{\MH^2}{\cmt^2}\right) &=&
20+6\beta_{\Pt}^2+3\beta_{\Pt}(\beta_{\Pt}^2+1)\ln(x_{\Pt})
+3(1-\beta_{\Pt}^2)\ln^2(x_{\Pt}),
\nn\\
\tau_{\PH\PW\PW}\left(\frac{\MH^2}{\cmt^2}\right) &=&
8+12\beta_{\Pt}^2+3\beta_{\Pt}(3\beta_{\Pt}^2-1)\ln(x_{\Pt})
+\frac{3}{2}(1-\beta_{\Pt}^2)^2\ln^2(x_{\Pt}),
\eeqar
where
\beq
\beta_{\Pt} = \sqrt{1-\frac{4\cmt^2}{\MH^2}}, \qquad
x_{\Pt} = \frac{\beta_{\Pt}-1}{\beta_{\Pt}+1}.
\eeq
They have the property $\tau_{\PH\PV\PV}(0)=0$, i.e.\ they quantify
the deviation of the $\Mt^2/\MW^2$-enhanced corrections from the pure
heavy-top limit, which are made explicit in \refeq{eq:IBA}.  Note that
we consistently use the complex top mass $\cmt$ instead of $\Mt$.  The
correction factor $\delta_{\Coul}$ containing the Coulomb singularity
reads \cite{Fadin:1993kg}
\beqar
\delta_{\Coul}(s,k_+^2,k_-^2) &=& \frac{\alpha(0)}{\bar\beta}
\Im\left\{\ln\left(\frac{\beta-\bar\beta+\Delta_M}
        {\beta+\bar\beta+\Delta_M}\right)\right\},\nl
\bar\beta &=& \frac{\sqrt{s^2+k_+^4+k_-^4-2sk_+^2-2sk_-^2-2k_+^2k_-^2}}{s},\nl
\beta &=&  \sqrt{1-\frac{4\cmw^2}{s}}, \qquad
\Delta_M = \frac{|k_+^2-k_-^2|}{s},
\eeqar
with the fine-structure constant $\alpha(0)$.
The auxiliary function 
\beq
g(\bar\beta) = \left(1-\bar\beta^2\right)^2
\eeq
restricts the impact of $\delta_{\Coul}$ to the WW~threshold region where 
it is valid.
The QCD corrections $\de_{\PV\to f_i\bar f_j}^{\QCD}$
to the gauge-boson decays are given by
\beq
\de_{\PV\to l_i\bar l_j}^{\QCD} = 0, \qquad
\de_{\PV\to q_i\bar q_j}^{\QCD} = 
\frac{\alpha_{\mathrm{s}}}{\pi},
\eeq
according to whether $f_i \bar f_j$ is a lepton
($l_i\bar l_j$) or a quark ($q_i\bar q_j$) pair, respectively.
Finally, the ``constants'' $c_{\PH\PV\PV}$ have been introduced to
account for a sizeable, but widely $\MH$-independent offset in the relative
corrections to the $\PH\PV\PV$ vertices
that is induced by non-leading corrections.
In practice, it is often sufficient to set $c_{\PH\PV\PV}$ to a
numerical constant, which can be determined from a comparison with
the full correction to the $\PH\to\PV\PV\to4f$ process.
For the input parameters given in \refse{se:input} we find
that the choice
\beq
c_{\PH\PZ\PZ} = 3\%, \qquad
c_{\PH\PW\PW} = 4\%
\eeq
is appropriate.  The values of $c_{\PH\PV\PV}$ will certainly not
change significantly if the input parameters of \refse{se:input} vary
within their experimental uncertainties.

When defining the IBA for the final states $f\bar f f\bar f$ and 
$f\bar f f'\bar f'$ we make use of the fact that the two 
subamplitudes in \refeq{eq:zzsym} and \refeq{eq:mixed} have
a very small interference. Therefore, we define the IBA for
the squared matrix element from the IBA of the corresponding
squared subamplitudes and take into account the interference
in lowest order without modification.

In deriving the contributions enhanced by $\Mt^2$ in (6.1), we used a
double-pole approximation and implicitly assumed an integration over
all decay angles. Without this assumption, corrections to angular
correlations are induced by the interference of the two relevant
formfactors describing the $\PH\PV\PV$ vertices. Hence, the IBA (6.1)
is in first place constructed for partial widths.

\section{Numerical results}
\label{se:numerics}

\subsection{Input parameters and setup}
\label{se:input}

We use the following set of input parameters \cite{Eidelman:2004wy},
\beqar
\begin{array}[b]{r@{\,}lr@{\,}lr@{\,}l}
G_{\mu} &= 1.16637\times 10^{-5}\GeV^{-2}, \qquad &
\al(0) &= 1/137.03599911, \qquad & \alpha_{\mathrm{s}} &= 0.1187,\\
\MW^{\LEP} &= 80.425\GeV, & \GW^{\LEP} &= 2.124\GeV, && \\
\MZ^{\LEP} &= 91.1876\GeV,& \GZ^{\LEP} &= 2.4952\GeV, && \\
\Me &= 0.51099892 \MeV, & \Mmy &= 105.658369 \MeV, 
&\Mta &= 1.77699\GeV,\\
\Mu &= 66 \MeV, & \Mc &=1.2 \GeV, & \Mt &= 174.3\GeV, \\
\Md &= 66 \MeV, & \Ms &=150 \MeV, & \Mb &= 4.3\GeV.
\end{array}
\eeqar

The masses of the light quarks are adjusted to reproduce the hadronic
contribution to the photonic vacuum polarization of
\citere{Jegerlehner:2001ca}. As discussed in \refse{se:inputscheme},
we use the $\GF$ scheme, \ie we derive the electromagnetic coupling
constant from the Fermi constant according to \refeq{eq:al-GF}.
We use $\alpha_{\GF}$ everywhere except for the couplings of the
collinear photons, as described in \refse{se:fsr}. In this case we use
$\al(0)$, because this reflects the coupling behaviour of real
photons.

Using the complex-mass scheme, we employ a fixed width in the resonant
W- and Z-boson propagators in contrast to the approach used at LEP to
fit the W~and Z~resonances, where running widths are taken.
Therefore, we have to convert the ``on-shell'' values of $M_V^{\LEP}$
and $\Ga_V^{\LEP}$ ($V=\PW,\PZ$), resulting from LEP, to the ``pole
values'' denoted by $M_V$ and $\Ga_V$ in this paper. The relation between
the two sets of values is given by \cite{Bardin:1988xt}
\beq\label{eq:m_ga_pole}
M_V = M_V^{\LEP}/
\sqrt{1+(\Ga_V^{\LEP}/M_V^{\LEP})^2},
\qquad
\Ga_V = \Ga_V^{\LEP}/
\sqrt{1+(\Ga_V^{\LEP}/M_V^{\LEP})^2},
\eeq
leading to
\beqar
\begin{array}[b]{r@{\,}l@{\qquad}r@{\,}l}
\MW &= 80.397\GeV, & 
\MZ &= 91.1535\GeV.
\label{eq:m_ga_pole_num}
\end{array}
\eeqar
We make use of these mass parameters in the numerics discussed below,
although the difference between using $M_V$ or $M_V^{\LEP}$ would be
hardly visible.  The widths of the gauge bosons W and Z, $\GW$ and
$\GZ$, are calculated from the above input including $\Oa$
corrections, but using real mass parameters everywhere.
Alternatively, the experimental widths calculated from
\refeq{eq:m_ga_pole} could be used, but the procedure pursued here
ensures that the ``effective branching ratios'' of the W's and Z's,
which result from the integration over their decays, add up to one if
all decay channels are summed over.  The gauge-boson widths depend on
the Higgs mass only weakly.  For the Higgs masses
$\MH=140,170,200\GeV$ the corresponding values are given in
\refta{tab:width}.  These values are used everywhere in the complex
masses, \ie we also
apply the $\Oa$-corrected W and Z~widths for the lowest-order
predictions.

In order to improve the corrections for partial Higgs decay widths for
a Higgs mass near the $\Pt\bar\Pt$ threshold ($\MH\sim2\Mt$), we
evaluate loop diagrams with internal top quarks with a complex
top-quark mass $\cmt=\sqrt{\Mt^2-\ri\Mt\Gt}$. To this end, we set the
top-quark width to its lowest-order prediction in the SM,
\beq
\Gt = \frac{\GF(\Mt^2-\MW^2)^2(\Mt^2+2\MW^2)}{8\pi\sqrt{2}\Mt^3}
= 1.54\GeV,
\eeq
where the simplifications $\Mb=0$ and $V_{\Pt\Pb}=1$ are used.

The angular distributions in \refse{se:angdistr} are defined in the
rest frame of the Higgs boson.  All observables are calculated without
applying phase-space cuts, and, if not stated otherwise, a photon
recombination is performed.  More precisely, if the invariant mass of
a photon and a charged fermion is smaller than $5\GeV$, the photon
momentum is added to the fermion momentum in the histograms.  If this
condition applies to more than one fermion the photon is recombined
with the fermion that yields the smallest invariant mass for the
resulting fermion--photon pair.

All but the lowest-order predictions contain the higher-order FSR as
described in \refse{se:fsr} as well as the two-loop corrections
proportional to $\GF^2\MH^4$ given in \refse{se:gf2mh4}.

\subsection{Results for partial decay widths}

In \refta{tab:width} the partial decay width including $\Oa$
corrections is shown for different decay channels and different values
of the Higgs mass.
\begin{table}
\centerline{
\begin{tabular}{|c|c|c|c|c|c|c|c|}
\hline
& $\MH[\GeV]$ & \multicolumn{2}{c|}{$140$} 
 & \multicolumn{2}{c|}{$170$} 
 & \multicolumn{2}{c|}{$200$} 
 \\
\hline
\hline
& $\GW[\GeV]$ & \multicolumn{2}{c|}{$2.09052...$} 
 & \multicolumn{2}{c|}{$2.09054...$} 
 & \multicolumn{2}{c|}{$2.09055...$} 
 \\
\hline
& $\GZ[\GeV]$ & \multicolumn{2}{c|}{$2.50278...$} 
 & \multicolumn{2}{c|}{$2.50287...$} 
 & \multicolumn{2}{c|}{$2.50292...$} 
 \\
\hline
\hline
$\PH\;\to$& & $\Gamma [\MeV]$ & $\delta[\%]$ & 
$\Gamma [\MeV]$ & $\delta[\%]$ & $\Gamma [\MeV]$ & $\delta[\%]$\\
\hline
$\mathrm{e^- e^+}\mu^-\mu^+$                                     
 & corrected
 &    0.0012628(5)
 & 2.3
 &    0.020162(7)
 & 2.7
 &    0.8202(2)
 & 4.4
  \\
 &
 lowest order
 &    0.0012349(4)
 &
 &    0.019624(5)
 &
 &    0.78547(8)
 &
  \\\hline
${l^- l^+}{l^- l^+}$                               
 & corrected
 &    0.0006692(2)
 & 2.1
 &    0.010346(3)
 & 2.7
 &    0.41019(8)
 & 4.4
  \\
 $l=\Pe,\mu$ &
 lowest order
 &    0.0006555(2)
 &
 &    0.010074(2)
 &
 &    0.39286(4)
 &
  \\\hline
$\nu_\mathrm{e}\mathrm{e^+}\mu^-\bar\nu_{\mu}$                   
 & corrected
 &    0.04807(2)
 & 3.7
 &    4.3109(9)
 & 6.2
 &   12.499(3)
 & 5.0
  \\
 &
 lowest order
 &    0.04638(1)
 &
 &    4.0610(7)
 &
 &   11.907(2)
 &
  \\\hline
$\nu_l {l^+} {l^-}\bar\nu_{l}$     
 & corrected
 &    0.04914(2)
 & 3.7
 &    4.344(1)
 & 6.1
 &   14.133(3)
 & 5.0
  \\
 $l=\Pe,\mu$ &
 lowest order
 &    0.04738(2)
 &
 &    4.0926(8)
 &
 &   13.458(2)
 &
  \\\hline
\end{tabular} }
\caption{Partial decay widths for $\PH\to4\,$leptons including 
\Oa{} and ${\cal O}(\GF^2\MH^4)$ corrections and corresponding
relative corrections for various  decay channels and different Higgs
masses.} 
\label{tab:width}
\end{table}
In parentheses the statistical error of the phase-space integration is
indicated, and $\delta=\Ga/\Ga_0-1$ labels the relative corrections.
The first two channels, $\mathrm{e^- e^+}\mu^-\mu^+$ and ${l^-
  l^+}{l^- l^+}$, $l=\Pe,\mu$, result from the decay
$\PH\to\PZ\PZ\to4f$. The partial widths for only electrons or muons in
the final state are equal in the limit of vanishing external fermion
masses, since for collinear-safe observables, such as the partial
widths, the fermion-mass logarithms cancel. The corresponding
lowest-order matrix elements are given in \refeq{eq:zz} and
\refeq{eq:zzsym}, respectively.  The width for $\PH\to{l^- l^+}{l^-
  l^+}$ is typically smaller by a factor 2, because it gets a factor
$1/4$ for identical particles in the final state and it proceeds in
lowest order via two Feynman diagrams that are related by the exchange
of two outgoing electrons and that have only a small interference.
The channel $\nu_\mathrm{e}\mathrm{e^+}\mu^-\bar\nu_{\mu}$
\refeq{eq:ww} results from the decay $\PH\to\PW\PW\to4f$, while the
last channel
$\nu_l l^+ l^-\bar\nu_l$
\refeq{eq:mixed} receives contributions from both
the decay into W and into Z~bosons.  The larger the Higgs mass, the
larger is the decay width, because the available phase space grows.

\begin{figure}
\setlength{\unitlength}{1cm}
\centerline{
\begin{picture}(7.7,8)
\put(-1.7,-14.5){\includegraphics{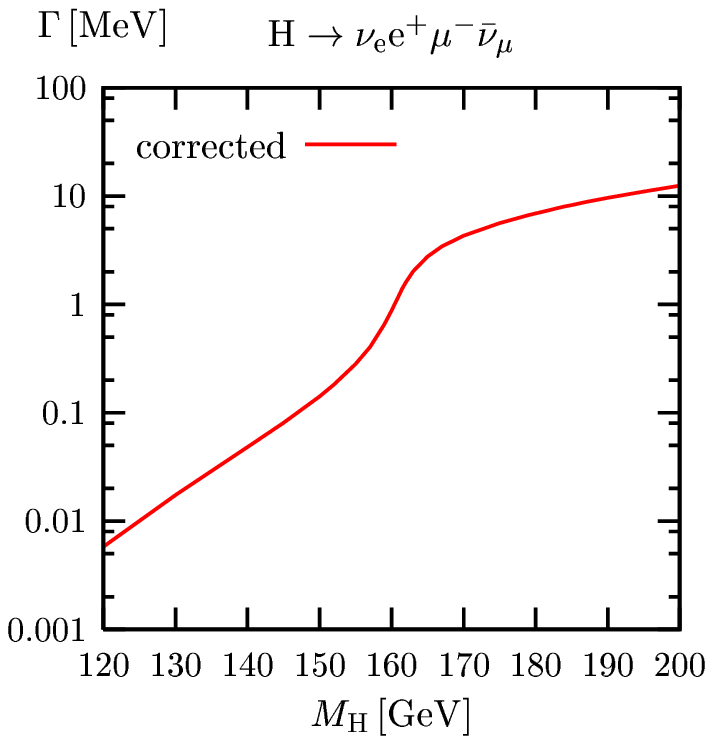}}
\end{picture}
\begin{picture}(7.5,8)
\put(-1.7,-14.5){\includegraphics{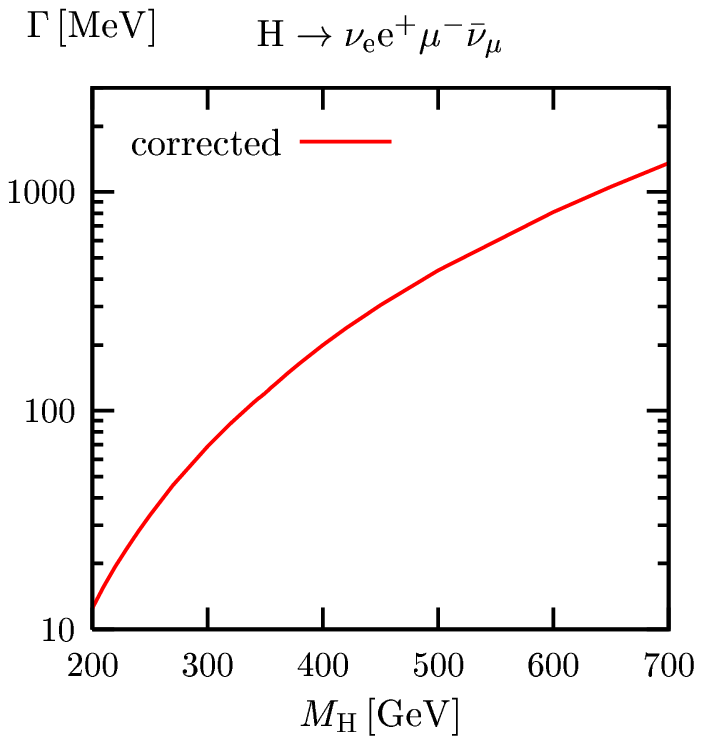}}
\end{picture} }
\centerline{
\begin{picture}(7.7,8)
\put(-1.7,-14.5){\includegraphics{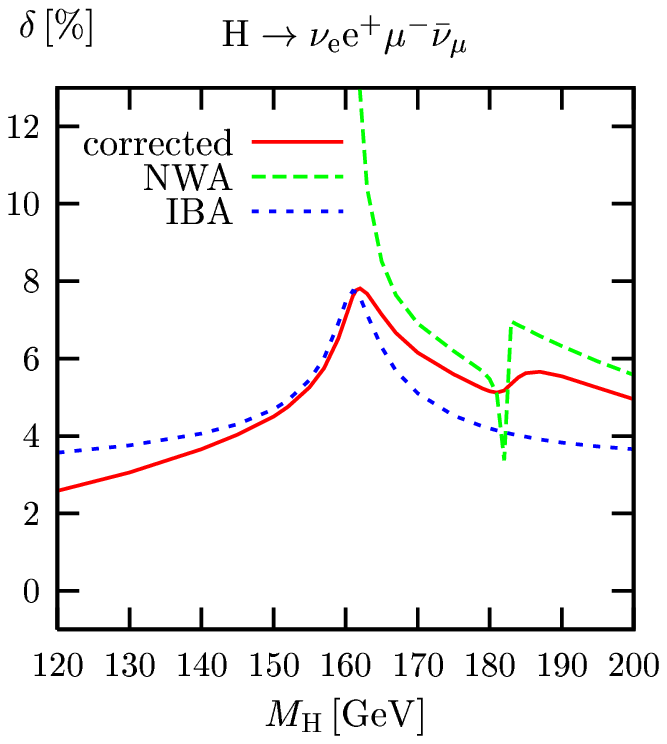}}
\end{picture}
\begin{picture}(7.5,8)
\put(-1.7,-14.5){\includegraphics{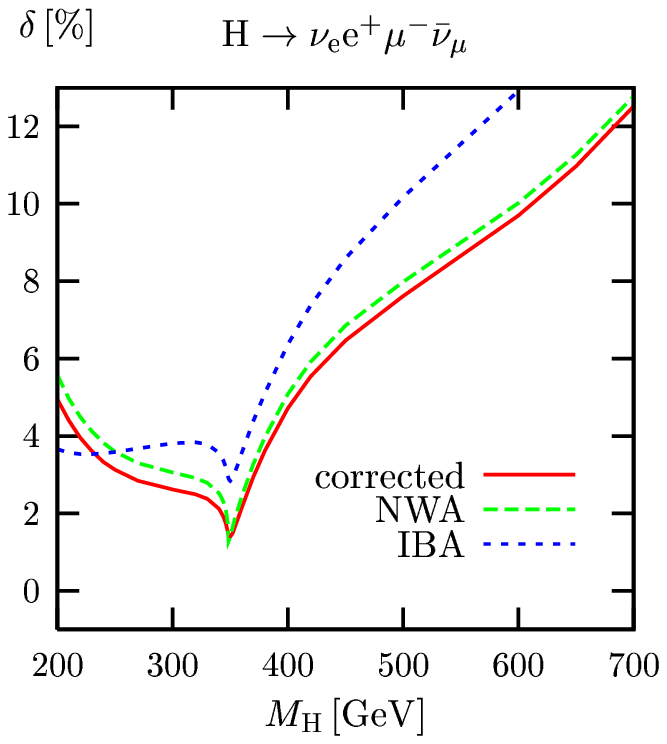}}
\end{picture} }
\caption{Partial decay width for $\PH\to\Pne\Pep\Pmum\Pnmubar$ 
  as a function of the Higgs mass. The upper plots show the absolute
  prediction including $\Oa$ and ${\cal O}(\GF^2\MH^4)$ corrections,
  and the lower plots show the comparison of the corresponding
  relative corrections with the NWA and IBA.}
\label{fig:sqrtsww}
\end{figure}

In the two upper plots of \reffi{fig:sqrtsww} we show the partial 
decay width for the final state $\Pne\Pep\Pmum\Pnmubar$ as 
a function of the Higgs mass. 
The lower plots show the corrections relative to the lowest-order
result. As already explained, we always normalize to the lowest-order
result that already includes the $\Oa$-corrected gauge-boson width in
the complex masses of the gauge bosons.  A large fraction of the $\Oa$
corrections is transfered to the lowest-order decay width by applying
the $\GF$ scheme. Thus, the corrections are at the order of 2--8\% for
moderate Higgs masses.  However, for large Higgs masses the
corrections become larger and reach about 13\% at $\MH=700\GeV$. In
this region the leading two-loop corrections already amount to about
4\%.  Around the $\PW\PW$ threshold at $160\GeV$ the Coulomb
singularity, which originates from soft-photon exchange between the
two slowly moving W~bosons, is reflected in the shape of the curve.
The influence of diagrams with a Higgs boson splitting into a virtual
Z-boson pair (ZZ~threshold) is visible at $\MH\sim2\MZ$.  At about
$2\Mt$ the $\Pt\Ptbar$ threshold is visible.

For stable W or Z~bosons, \ie in the limit $\Ga_V\to 0$ ($V=\PW,\PZ$),
it is possible to define a narrow-width approximation (NWA) where the
matrix elements factorize into the decay $\PH\to\PV\PV$ and the
subsequent decay of the gauge bosons into fermions. By definition the
NWA is only applicable above the WW or ZZ~threshold. However, its
analytical structure and evaluation is considerably simpler than in
the case of the full decay $\PH\to\PW\PW/\PZ\PZ\to4f$ with off-shell
gauge bosons.  Therefore, above threshold the NWA allows for an
economic way of calculating relative $\Oa$ corrections to the
integrated decay width, while the lowest-order contribution may, of
course, still take into account unstable gauge bosons.  Following this
line of thought, we define
\beq
\Ga^{\NWA} = \Ga_0 \, \frac{\Ga^{\NWA}_1} {\Ga^{\NWA}_0},
\eeq
with
\beq
\Ga^{\NWA}_1 = \Ga_{\PH\PV\PV,1}\, \frac{\Ga_{\PV f_1\bar f_2,1}
\Ga_{\PV f_3\bar f_4,1}}{\Ga_{\PV,1}\Ga_{\PV,1}},
\eeq
and
\beq
\Ga^{\NWA}_0 = \Ga_{\PH\PV\PV,0}\, \frac{\Ga_{\PV f_1\bar f_2,0}
\Ga_{\PV f_3\bar f_4,0}}{\Ga_{\PV,1}\Ga_{\PV,1}}.
\eeq
The indices ``0'' and ``1'' label lowest-order and $\Oa$-corrected
results, respectively.  The Higgs-mass-enhanced two-loop terms,
described in \refse{se:gf2mh4}, have also been included in
$\Ga_{\PH\PV\PV,1}$.  In order to be consistent we again use the
$\Oa$-corrected total width for the gauge bosons in the denominators
of the branching ratios in $\Ga^{\NWA}_0$.  We note that we have
rederived all necessary $\Oa$ corrections entering the NWA; the
hard-photon corrections to the decay $\PH\to\PW\PW$ have been
successfully checked against the expression given in
\citere{Kniehl:1991xe}.  The NWA is evaluated with real gauge-boson
and top-quark masses.

A few GeV above the corresponding gauge-boson-pair threshold the NWA
agrees with the complete corrections within 1\%.  Near
$\MH=2\MZ\sim180\GeV$ the loop-induced ZZ~threshold can be seen in the
relative corrections to $\PH\to\PW\PW\to\Pne\Pep\Pmum\Pnmubar$ shown
in \reffi{fig:sqrtsww}.  In the NWA this threshold leads to a
singularity visible as a sharp peak; in the off-shell calculation in
the complex-mass scheme this singular structure is smeared out,
because the finite Z-boson width is taken into account.  Since the ZZ
threshold corresponds to the situation where two Z~bosons become on
shell in the loop, the latter description with the singularity
regularized by a finite $\GZ$ is closer to physical reality.  An
analogous situation can be seen near $\MH=2\Mt\sim350\GeV$ for the
$\Pt\bar\Pt$ threshold with top quarks in the loops. Again the
inclusion of the top decay width $\Gt$, as done in the off-shell
calculation, yields the better description.  In \reffi{fig:sqrtsww} we
show also the relative corrections obtained in the IBA of
\refeq{eq:IBA}. The Coulomb singularity and the $\Pt\bar\Pt$ threshold
are well described, and the full corrections are reproduced within
$\lsim2\%$ for Higgs masses below $400\GeV$.

\begin{figure}
\setlength{\unitlength}{1cm}
\centerline{
\begin{picture}(7.7,8)
\put(-1.7,-14.5){\includegraphics{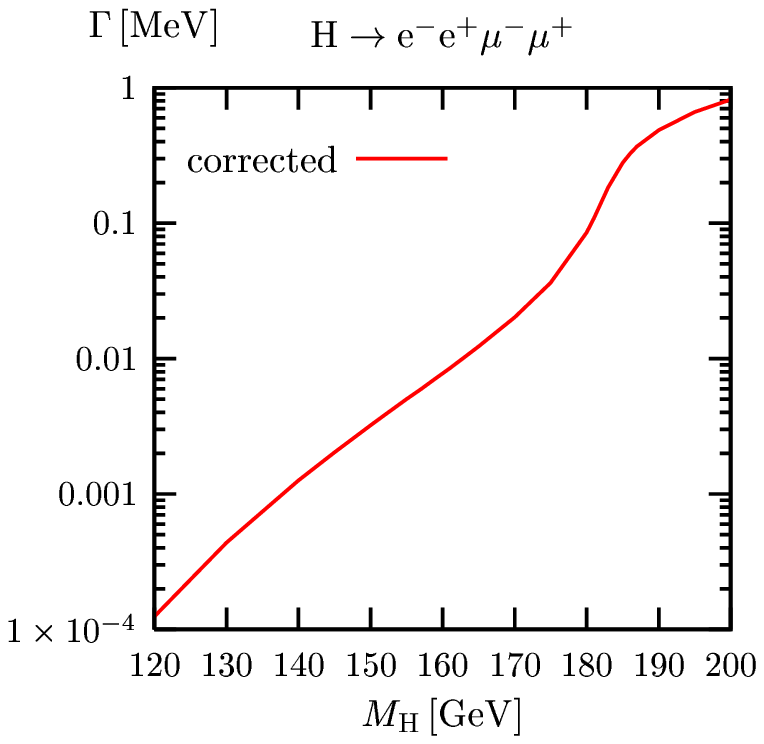}}
\end{picture}
\begin{picture}(7.5,8)
\put(-1.7,-14.5){\includegraphics{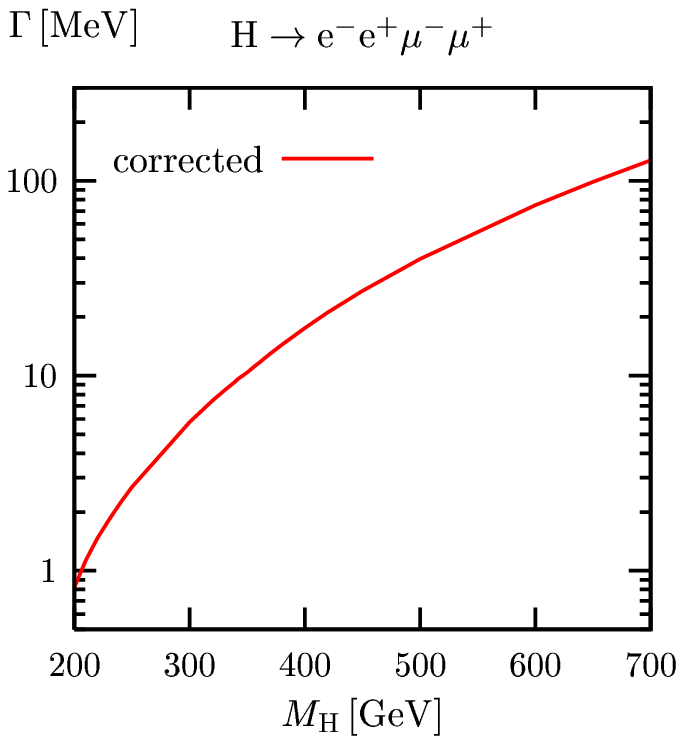}}
\end{picture} }
\centerline{
\begin{picture}(7.7,8)
\put(-1.7,-14.5){\includegraphics{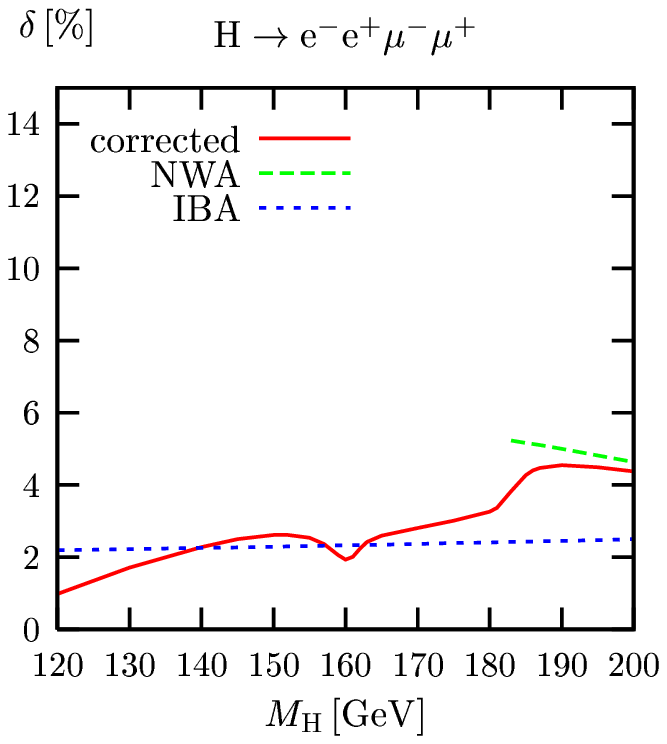}}
\end{picture}
\begin{picture}(7.5,8)
\put(-1.7,-14.5){\includegraphics{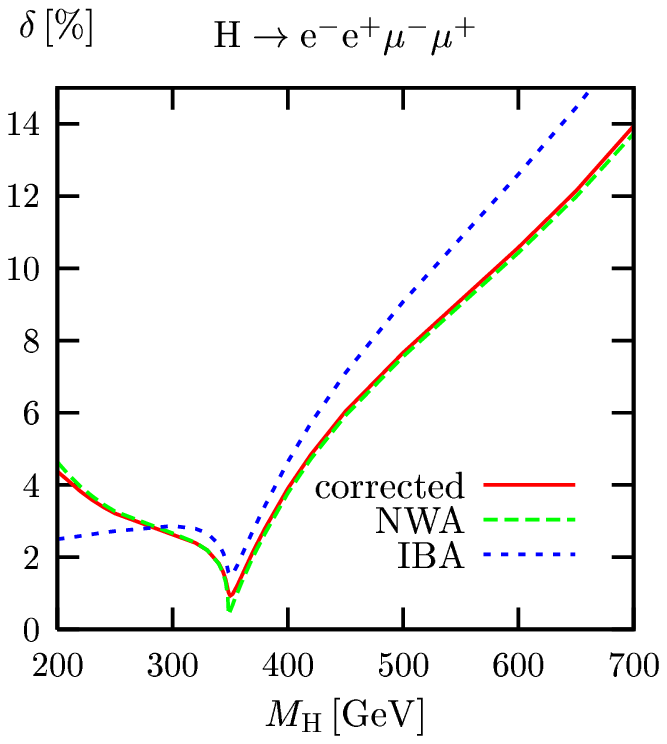}}
\end{picture} }
\caption{Partial decay width for $\PH\to\Pem\Pep\Pmum\Pmup$ 
  as a function of the Higgs mass. The upper plots show the absolute
  prediction including $\Oa$ and ${\cal O}(\GF^2\MH^4)$ corrections,
  and the lower plots show the comparison of the corresponding
  relative corrections with the NWA and IBA.}
\label{fig:sqrtszz}
\end{figure}

The plots in \reffi{fig:sqrtszz} show the decay width and the 
relative correction for the final state $\Pem\Pep\Pmum\Pmup$. 
The corrections are between 2\% and 4\% for moderate Higgs masses and
rise to more than 10\% for large Higgs masses.  At a Higgs mass of
about $160\GeV$ the influence of the $\PW\PW$ threshold can be
observed.  As explained above, the behaviour of the corrections as a
function of the Higgs mass is smooth, because the finite W-boson width
is also used in the loop integrals.  In contrast to the decay
$\PH\to\Pne\Pep\Pmum\Pnmubar$, there is no Coulomb singularity in this
channel because the Z~boson is electrically neutral.  The NWA
reproduces the complete result within $0.5\%$ not too close to the
threshold. The IBA agrees with the complete calculation to better than
2\% for not too large Higgs masses.

\subsection{Comparison of partial widths with \HDECAY}

Predictions for the partial decay widths of the Higgs boson can also
be obtained with various program packages, such as
\HDECAY~\cite{Djouadi:1997yw}.  \HDECAY{} contains the lowest-order
decay width for $\PH\to\PV^{(*)}\PV^{(*)}$ and the leading one-loop
corrections $\propto\GF\MH^2$ and two-loop corrections $\propto
\GF^2\MH^4$.  In order to obtain the decay width for
$\PH\to\PW\PW/\PZ\PZ\to 4f$, we define
\beq
\Ga^{\mathrm{HD}} = \Ga_{\PH\PV\PV}^{\mathrm{HD}} \,
\frac{\Ga_{\PV f_1f_2,0}}{\Ga_{\PV,1}} \,
\frac{\Ga_{\PV f_3f_4,0}}{\Ga_{\PV,1}},
\label{eq:HD}
\eeq
where $\Ga_{\PH\PV\PV}^{\mathrm{HD}}$ is the decay width from \HDECAY.
In \refeq{eq:HD} the branching ratios of the gauge bosons are
normalized in the same way (lowest order in the numerator, corrected
total width in the denominator) as the effective branching ratios of
our lowest-order predictions for the $\PH\to VV\to4f$ partial widths;
otherwise a comparison would not be conclusive.

\begin{figure}
\setlength{\unitlength}{1cm}
\centerline{
\begin{picture}(7.7,8)
\put(-1.7,-14.5){\includegraphics{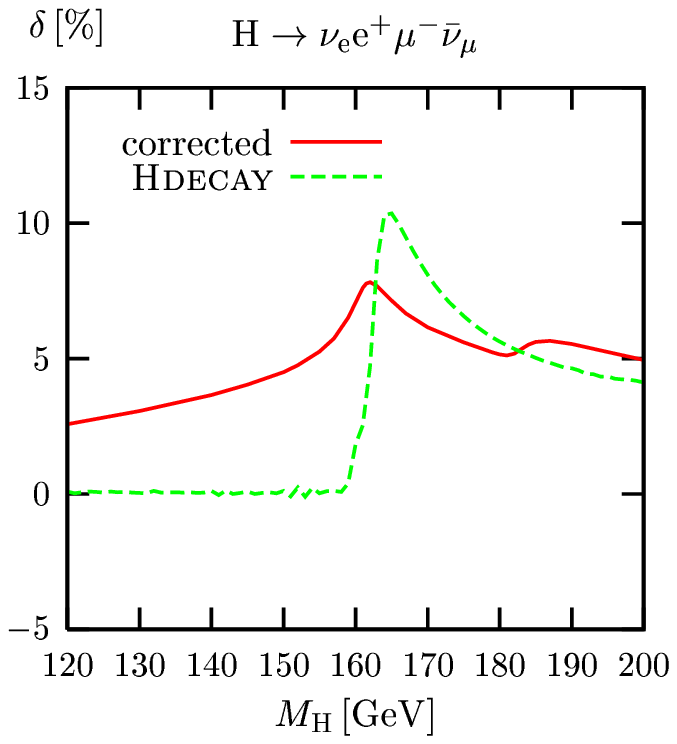}}
\end{picture}
\begin{picture}(7.5,8)
\put(-1.7,-14.5){\includegraphics{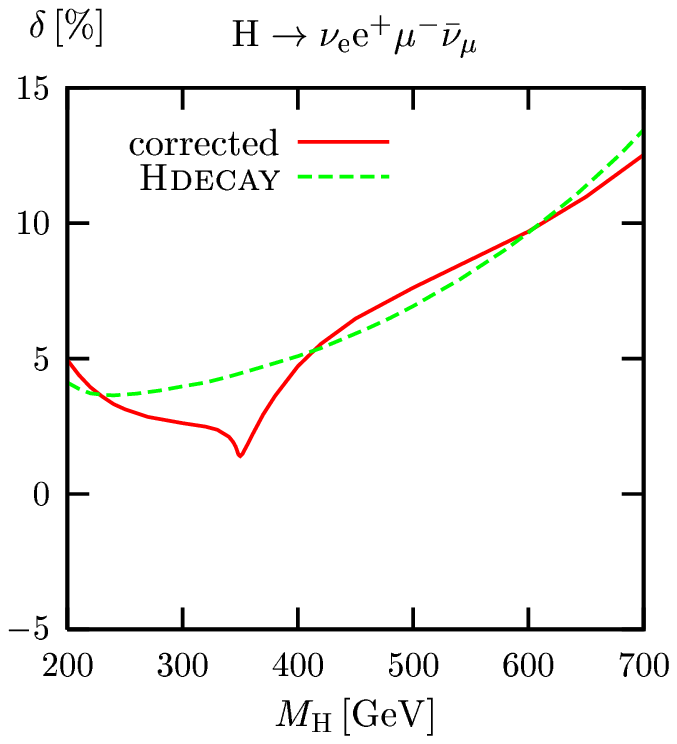}}
\end{picture} }
\centerline{
\begin{picture}(7.7,8)
\put(-1.7,-14.5){\includegraphics{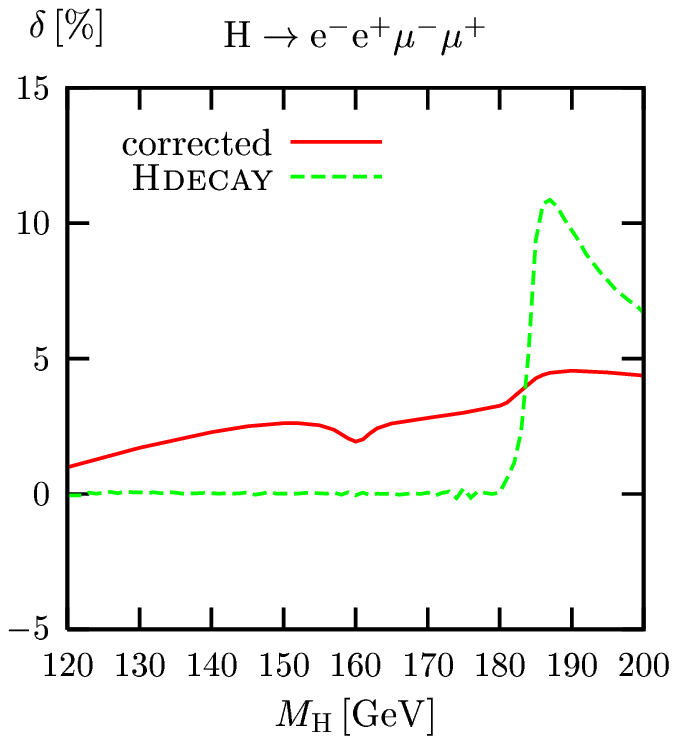}}
\end{picture}
\begin{picture}(7.5,8)
\put(-1.7,-14.5){\includegraphics{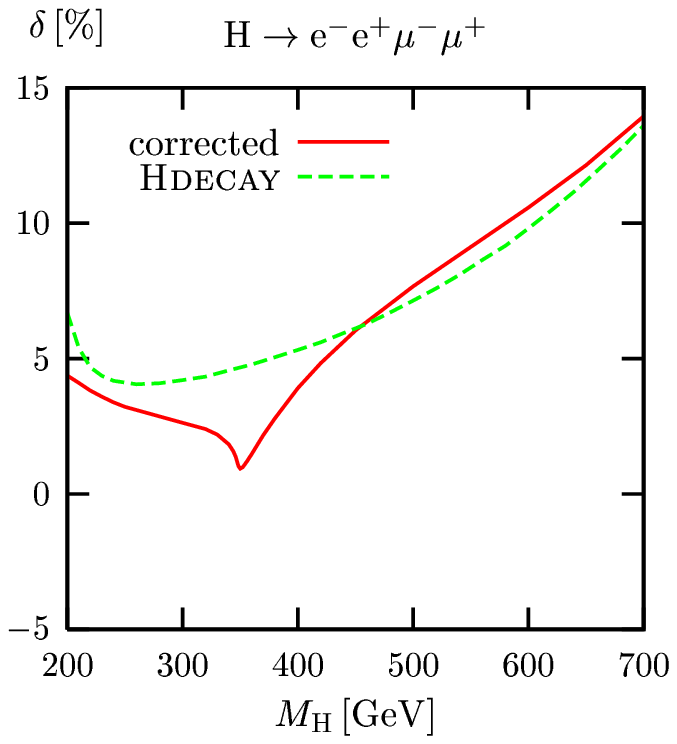}}
\end{picture} }
\caption{Predictions for the partial decay widths for
  $\PH\to\Pne\Pep\Pmum\Pnmubar$ and $\PH\to\Pem\Pep\Pmum\Pmup$
  obtained with the program \HDECAY\ normalized to the complete
  lowest-order decay width.  The corrections shown in
  \reffis{fig:sqrtsww} and \ref{fig:sqrtszz} are included for
  comparison.}
\label{fig:sqrts_hdec}
\end{figure}

The comparison in \reffi{fig:sqrts_hdec}, where $\Ga^{\mathrm{HD}}$ is
shown relative to our complete lowest-order prediction, shows that
\HDECAY{} agrees with our lowest-order prediction below the decay
threshold quite well.
In this region $\Ga_{\PH\PV\PV}^{\mathrm{HD}}$ consistently takes into
account the off-shell effects of the gauge bosons.  Above the
threshold \HDECAY\ neglects off-shell effects of the gauge bosons.
For large $\MH$ it follows our corrected result within a few per cent,
because the dominant radiative corrections $\propto\GF\MH^2$ and
$\propto\GF^2\MH^4$, which grow fast with increasing $\MH$, are
included in both calculations.  In the threshold region, off-shell
effects are, however, very important.  Here, the difference between
the complete off-shell result and the Higgs width for on-shell gauge
bosons amounts to more than 10\%.  In detail, \HDECAY{} interpolates
between the off-shell and on-shell results within a window of
$\pm2\GeV$ around threshold.  The maxima in the \HDECAY\ curves near
the WW and ZZ~thresholds in the upper and lower left plots of
\reffi{fig:sqrts_hdec}, respectively, are artifacts originating from
the on-shell phase space of the W or Z~bosons above threshold.
Approaching the threshold from above, the on-shell phase space, and
thus the corresponding partial decay width, tends to zero.  This
feature is avoided in \HDECAY\ by the interpolation.  The described
maxima in the \HDECAY\ curves have nothing to do with the maximum of
the correction near the WW~threshold in the upper left plot, which is
due to the Coulomb singularity.

\subsection{Invariant-mass distributions}
\label{se:invmassdistr}

\begin{figure}
\setlength{\unitlength}{1cm}
\centerline{
\begin{picture}(7.7,8)
\put(-1.7,-14.5){\includegraphics{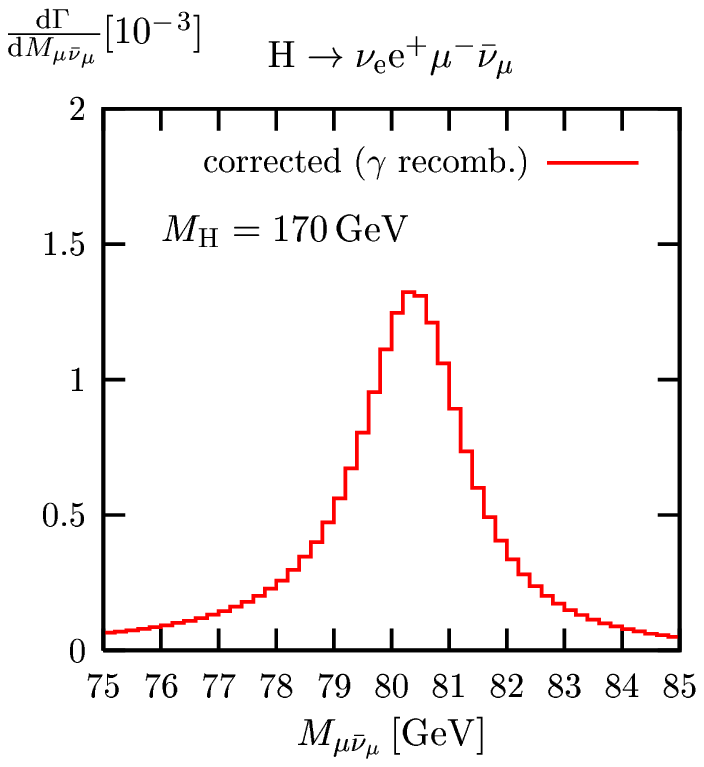}}
\end{picture}
\begin{picture}(7.5,8)
\put(-1.7,-14.5){\includegraphics{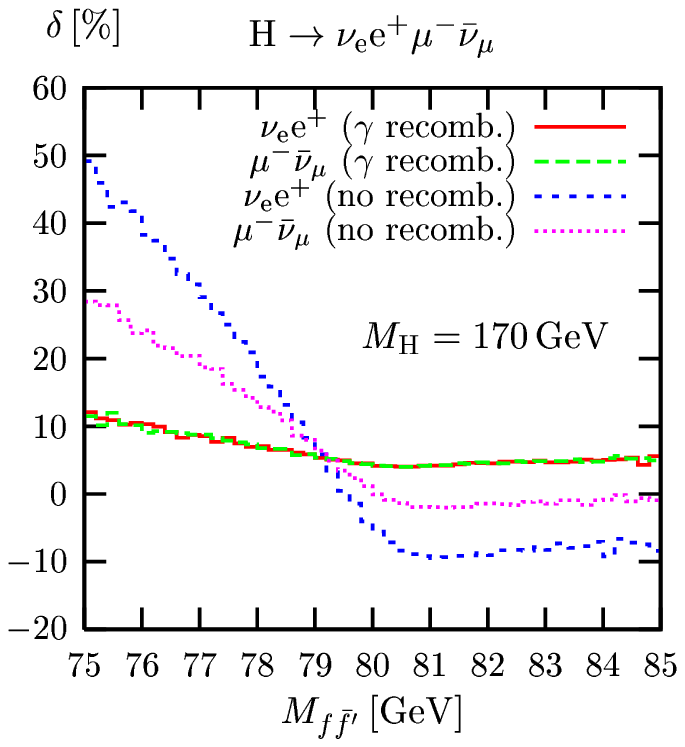}}
\end{picture} }
\centerline{
\begin{picture}(7.7,8)
\put(-1.7,-14.5){\includegraphics{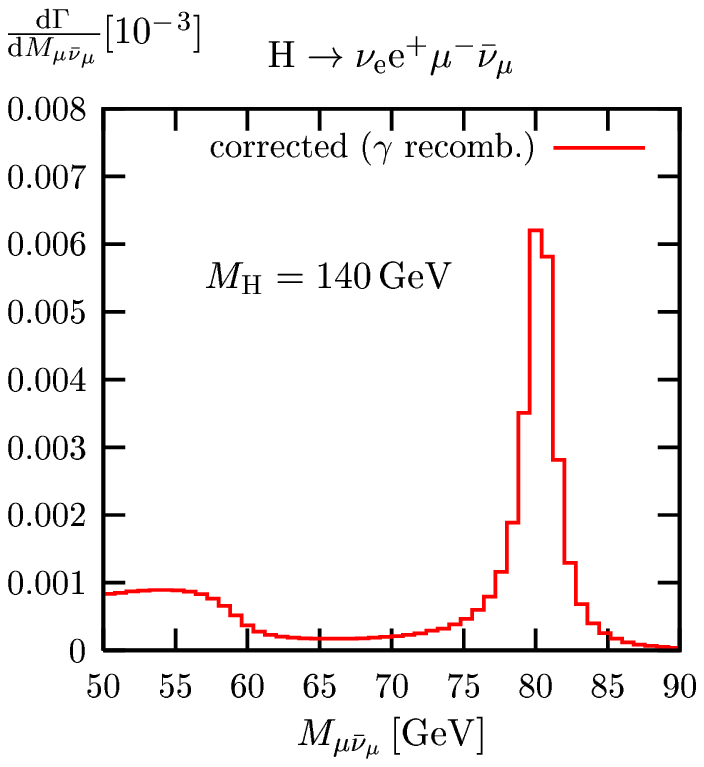}}
\end{picture}
\begin{picture}(7.5,8)
\put(-1.7,-14.5){\includegraphics{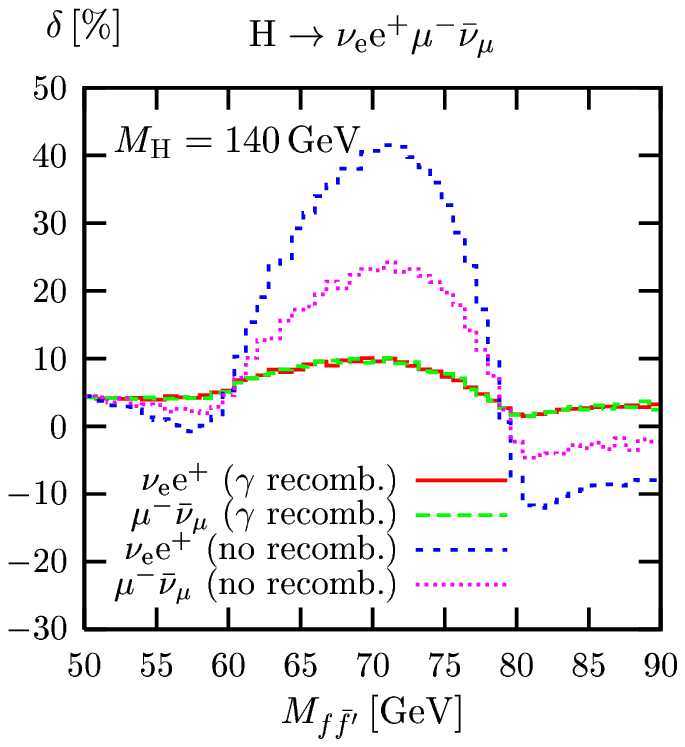}}
\end{picture} }
\caption{Distribution in the invariant mass of the $\Pmum\Pnmubar$
  (l.h.s.)\ pair and relative corrections to the distributions in the
  invariant masses of the $\Pne\Pep$ and $\Pmum\Pnmubar$ pairs
  (r.h.s.)\ in the decay $\PH\to\Pne\Pep\Pmum\Pnmubar$ for
  $\MH=170\GeV$ and $\MH=140\GeV$.}
\label{fig:winv}
\end{figure}

In \reffi{fig:winv} we study the invariant-mass distribution 
of the fermion pairs resulting from the decay of the W~bosons 
in the decay $\PH\to\Pne\Pep\Pmum\Pnmubar$. 
The plots on the l.h.s.\ show the distribution for $\Pmum\Pnmubar$
including corrections for $\MH=170\GeV$ and $\MH=140\GeV$, i.e.\ for
one value of $\MH$ above and one below the WW~threshold.  The plots on
the r.h.s.\ compare the relative corrections for $\Pne\Pep$ and
$\Pmum\Pnmubar$ both with and without photon recombination.  The
invariant mass $M_{f\bar f'}$ is calculated from the sum of the
momenta of the fermions $f$ and $f'$. If no photon recombination is
applied, always the bare momenta are taken. In the case of photon
recombination the momentum of recombined photons is included in the
invariant mass as described in \refse{se:input}.

For $\MH=170\GeV$, where the width is dominated by contributions where
both intermediate W~bosons are simultaneously resonant, the shape of
the curves in \reffi{fig:winv} can be understood as follows.  If one
of the fermions resulting from the decay of a resonant W~boson emits a
photon, the invariant mass $M_{f\bar f'}$ is reduced, giving rise to
an enhancement for small invariant masses.  Without photon
recombination these positive corrections are large due to the
appearance of logarithms of the small fermion masses. As the electron
mass is smaller, the corresponding logarithms yield a larger
contribution.  If photon recombination is applied, events are
rearranged from small invariant masses to large invariant masses.  In
this case, the observable is inclusive, i.e.\ the fermion mass
logarithms cancel owing to the KLN theorem, and the $\Pne\Pep$ and
$\Pmum\Pnmubar$ distributions do not differ.  The analogous phenomenon
has, e.g., been discussed for the related resonance processes
$\Pep\Pem\to\PW\PW\to 4\,$leptons
\cite{Denner:2000bj,Beenakker:1998gr} and $\ga\ga\to\PW\PW\to 4f$
\cite{Bredenstein:2005zk}.

For $\MH=140\GeV$, \ie below the threshold, only one W~boson can
become on shell. Thus, there is still a resonance around $M_{f\bar
  f'}\sim\MW$, but also an enhancement below an invariant mass of
$\MH-\MW\sim60\GeV$, where the other decaying W~boson can become
resonant.  Near the resonance at $M_{f\bar f'}\sim\MW$ the corrections
look similar to the doubly-resonant case discussed for $\MH=170\GeV$
above.  The same redistribution of events from higher to lower
invariant mass due to FSR happens as explained above.  Between $\MW$
and $\MH-\MW$ none of the W~bosons is resonant, and the contribution
to the lowest-order width is small. Therefore, owing to the
redistribution of events via photon emission the relative corrections
are large in this regime.  Below $\MH-\MW$, where the other W~boson
can become resonant, qualitatively the same FSR effects are visible as
in the vicinity of the resonance at $\MW$: apart from a constant
positive off-set in the relative corrections, events are distributed
from the right to the left of the maximum.

\begin{figure}
\setlength{\unitlength}{1cm}
\centerline{
\begin{picture}(7.7,8)
\put(-1.7,-14.5){\includegraphics{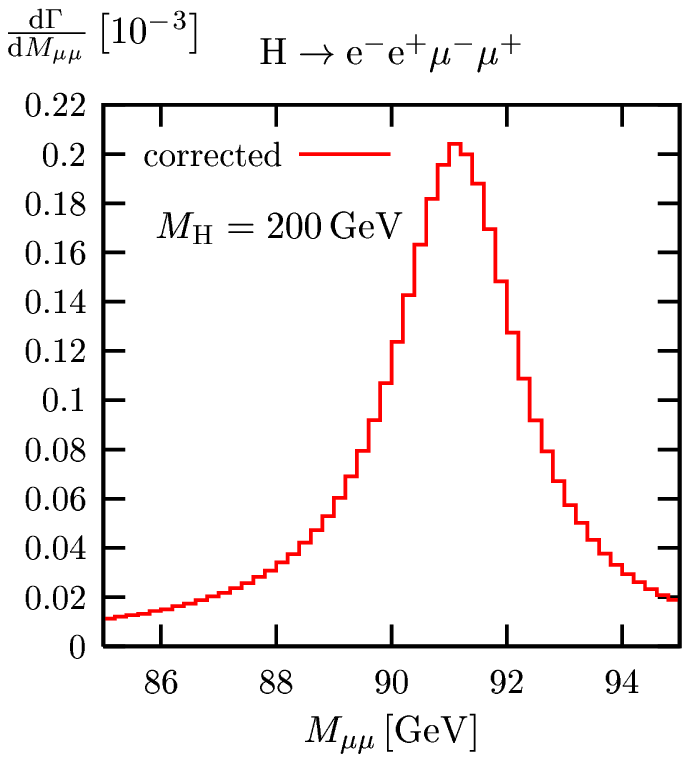}}
\end{picture}
\begin{picture}(7.5,8)
\put(-1.7,-14.5){\includegraphics{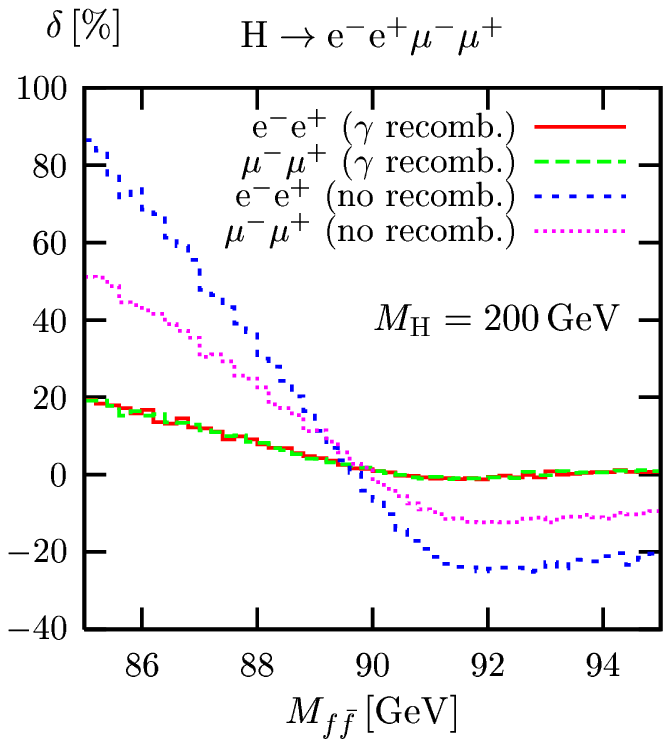}}
\end{picture} }
\centerline{
\begin{picture}(7.7,8)
\put(-1.7,-14.5){\includegraphics{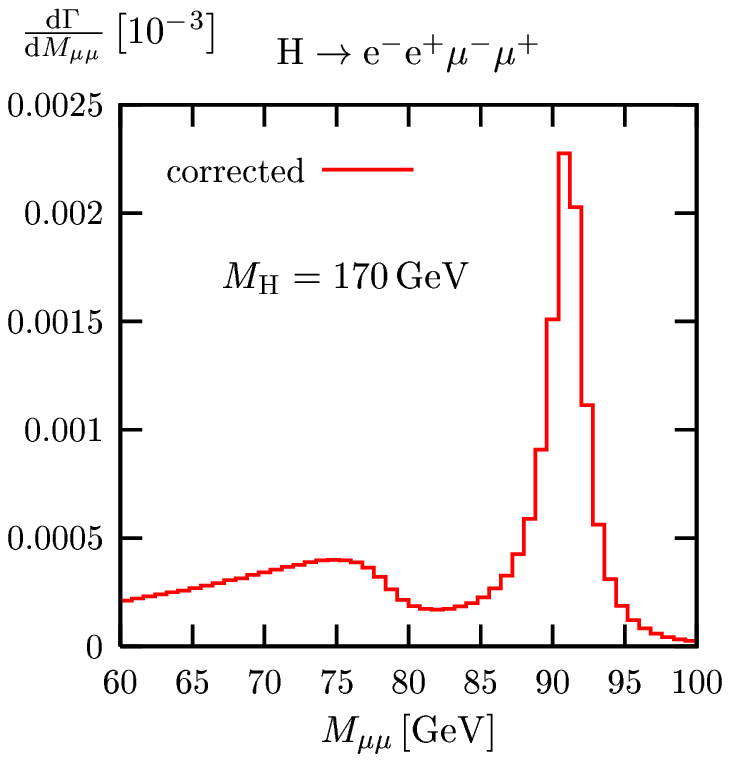}}
\end{picture}
\begin{picture}(7.5,8)
\put(-1.7,-14.5){\includegraphics{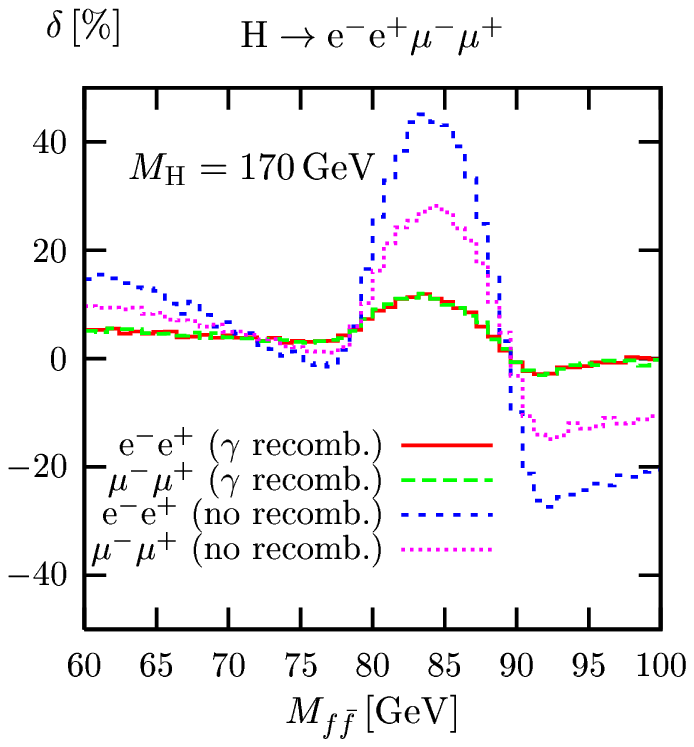}}
\end{picture} }
\caption{Distribution in the invariant mass of the $\Pmum\Pmup$ pair 
  (l.h.s.)\  and relative corrections to the distributions in the
  invariant masses of the $\Pem\Pep$ and $\Pmum\Pmup$ pairs (r.h.s.)\
  in the decay $\PH\to\Pem\Pep\Pmum\Pmup$ for $\MH=200\GeV$ and
  $\MH=170\GeV$.}
\label{fig:zinv}
\end{figure}

Figure~\ref{fig:zinv} shows the corresponding invariant-mass distributions 
for the decay $\PH\to\Pem\Pep\Pmum\Pmup$ with $\MH=200\GeV$ and $\MH=170\GeV$. 
The generic features of the plots are similar to the decay into
W~bosons.  For $\MH=200\GeV$, \ie above the ZZ~threshold, there is a
resonance region around $\MZ$, and the corrections become large in the
non-collinear-safe case.  Photon recombination rearranges the events,
so that the fermion logarithms cancel.  For Higgs masses below the
ZZ~threshold, such as for $\MH=170\GeV$, one Z~boson or the other is
resonant for $M_{f\bar f}\sim\MZ$ or $M_{f\bar f}\lsim \MH-\MZ$,
respectively. The shape and the large size of the corrections are due
to collinear FSR as explained above.  In \citere{Choi:2002jk} it was
pointed out that the kinematical threshold near $\MH-\MZ$ where the
other Z~boson can become on shell, which is at $M_{f\bar f}\sim80\GeV$
in \reffi{fig:zinv}, can be used to verify the spin of the Higgs
boson. While the rise of the width near this threshold is proportional
to $\be\propto\sqrt{[1-(\MZ+M_{f\bar f})^2/\MH^2][1-(\MZ-M_{f\bar
    f})^2/\MH^2]}$ for a spin-0 particle, it would be proportional to
$\be^3$ for a spin-1 particle.  Figure~\ref{fig:zinv} shows that the
radiative corrections influence the slope at the kinematical
threshold.

\begin{figure}
\setlength{\unitlength}{1cm}
\centerline{
\begin{picture}(7.7,8)
\put(-1.7,-14.5){\includegraphics{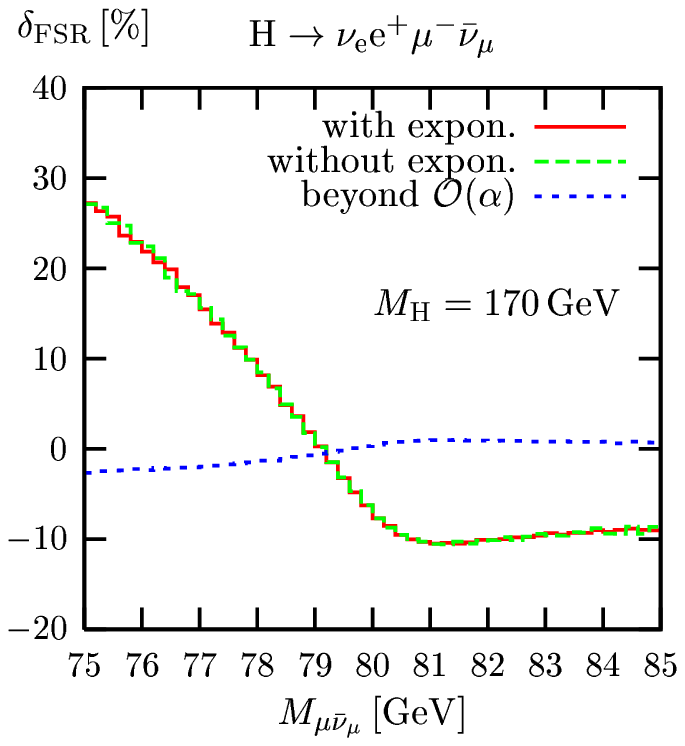}}
\end{picture}
\begin{picture}(7.5,8)
\put(-1.7,-14.5){\includegraphics{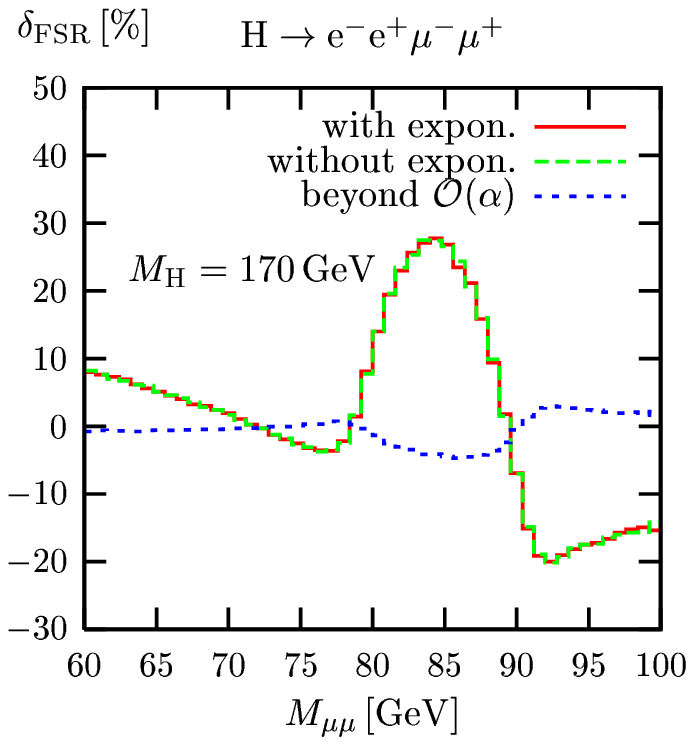}}
\end{picture} }
\caption{Influence of the leading logarithmic terms of FSR
on the invariant-mass distribution of $\Pmum\bar\Pnmu$ and 
$\Pmum\Pmup$ in the decays $\PH\to\Pne\Pep\Pmum\Pnmubar$ 
and $\PH\to\Pem\Pep\Pmum\Pmup$. The different curves correspond to 
the result with exponentiation, without exponentiation, and 
to the sum of $\al^2$ and $\al^3$ terms, which are labelled ``beyond $\Oa$''.}
\label{fig:fsr}
\end{figure}

Finally, in \reffi{fig:fsr} we investigate the influence of the
contribution $\de_{\mathrm{FSR}}$ of higher-order FSR to the complete
relative correction $\de$ on the invariant-mass distribution of
$\Pmum\bar\Pnmu$ and $\Pmum\Pmup$ in the decays
$\PH\to\Pne\Pep\Pmum\Pnmubar$ and $\PH\to\Pem\Pep\Pmum\Pmup$. The
invariant mass is defined via the momenta of the fermions alone, i.e.\ 
without photon recombination.  If photon recombination was applied,
the leading logarithmic FSR corrections, as described in
\refse{se:fsr}, would vanish completely.  Subtracting the $\Oa$ terms
\refeq{eq:Oafsr} with \refeq{eq:Oasf} from \refeq{eq:FSR} with the
structure functions \refeq{eq:GammaFSR} yields the contribution that
is beyond $\Oa$. In \reffi{fig:fsr} the impact of this contribution is
studied revealing corrections of up to $4\%$ in regions where the
lowest-order result is relatively small.  Figure~\ref{fig:fsr} also
shows the comparison between the structure function with and without
the exponentiation of the soft-photon parts in \refeq{eq:GammaFSR} and
\refeq{eq:FSRreexpand}, respectively.  The difference is beyond
$\Oaaa$ and turns out to be tiny.

\subsection{Angular distributions}
\label{se:angdistr}

\begin{figure}
\setlength{\unitlength}{1cm}
\centerline{
\begin{picture}(7.7,8)
\put(-1.7,-14.5){\includegraphics{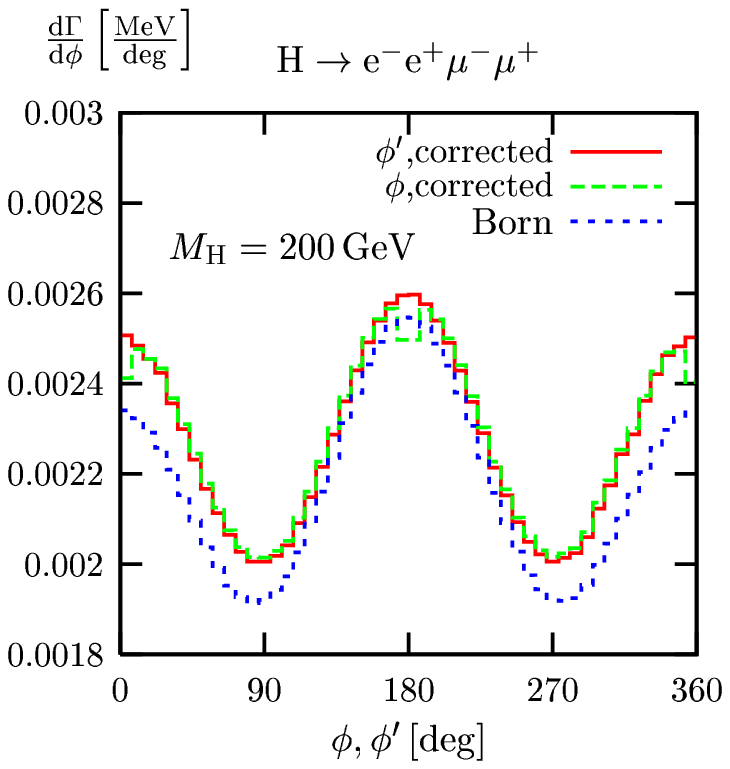}}
\end{picture}
\begin{picture}(7.5,8)
\put(-1.7,-14.5){\includegraphics{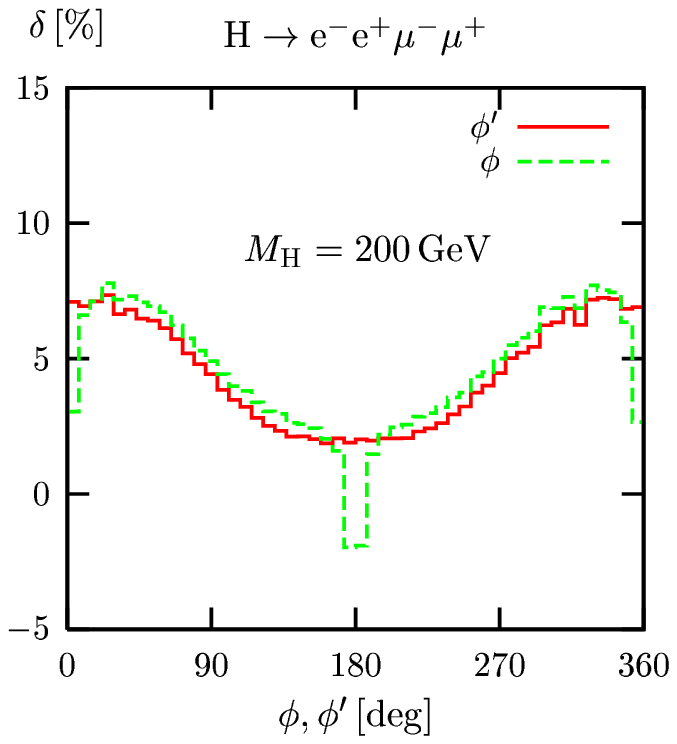}}
\end{picture} }
\caption{Distribution in the angle between the $\PZ\to l^-l^+$
  decay planes in the decay $\PH\to\Pem\Pep\Pmum\Pmup$ (l.h.s.)\ and
  corresponding relative corrections (with photon recombination)
  (r.h.s.)\ for $\MH=200\GeV$.}
\label{fig:phi}
\end{figure}

The investigation of angular correlations between the fermionic decay
products is an essential means of testing the properties of the Higgs
boson. In \citeres{Nelson:1986ki,Choi:2002jk} it was demonstrated how
the spin of the Higgs boson can be determined by looking at the angle
between the decay planes of the Z~bosons in the decay $\PH\to\PZ\PZ$.
This angle can be defined by
\beqar
\cos{\phi'} &=& 
\frac{({\bf k}_{12}\times{\bf k}_1)({\bf k}_{12}\times{\bf k}_3)}
     {|{\bf k}_{12}\times{\bf k}_1||{\bf k}_{12}\times{\bf k}_3|},
\nn\\
\sgn(\sin{\phi'}) &=& 
\sgn\{{\bf k}_{12}\cdot[({\bf k}_{12}\times{\bf k}_1)\times
                        ({\bf k}_{12}\times{\bf k}_3)]\},
\label{eq:phipr}
\eeqar
where ${\bf k}_{12}={\bf k}_1+{\bf k}_2$.  The l.h.s.\ of
\reffi{fig:phi} shows the differential decay width for
$\PH\to\Pem\Pep\Pmum\Pmup$ as a function of $\phi'$ revealing a
$\cos{2\phi'}$ term. As was noticed in
\citeres{Nelson:1986ki,Choi:2002jk}, this term would be proportional
to ($-\cos{2\phi'}$) if the Higgs boson was a pseudo-scalar.

Note that for non-photonic events 
the definition of $\phi'$ coincides with the definition given in 
\citere{Denner:2000bj} where ($-{\bf k}_{34}\times{\bf k}_3$) 
with ${\bf k}_{34}={\bf k}_3+{\bf k}_4$ was used instead of 
(${\bf k}_{12}\times{\bf k}_3$). 
Explicitly, $\phi$ was defined by
\beqar
\cos{\phi} &=& 
\frac{({\bf k}_{12}\times{\bf k}_1)(-{\bf k}_{34}\times{\bf k}_3)}
                {|{\bf k}_{12}\times{\bf k}_1||-{\bf k}_{34}\times{\bf k}_3|},
\nn\\
\sgn(\sin{\phi}) &=& 
\sgn\{{\bf k}_{12}\cdot[({\bf k}_{12}\times{\bf k}_1)\times
                           (-{\bf k}_{34}\times{\bf k}_3)]\}.
\label{eq:phi}
\eeqar
However, this definition yields large negative contributions at
$\phi=0^{\circ}$ and $\phi=180^{\circ}$. As was explained in
\citere{Denner:2000bj}, this is an effect of the suppressed phase
space in the real corrections. At $\phi=0^{\circ}$ and
$\phi=180^{\circ}$ the phase space for photonic events shrinks to the
configurations where the photon is either soft or lies in the decay
plane of the gauge bosons. Thus, the negative contributions from the
virtual corrections are not fully compensated by the real corrections.
Using ${\bf k}_{12}\times{\bf k}_3$ as in \refeq{eq:phipr} avoids this
suppression and gives rise to a smooth dependence of the corrections
on $\phi$ as can be seen on the r.h.s.\ of \reffi{fig:phi} which shows
the relative corrections for $\phi$ and $\phi'$ in the decay
$\PH\to\Pem\Pep\Pmum\Pmup$.  Since the difference of $\phi$ and
$\phi'$ is only due to photons, this, again, emphasizes the 
influence of the photon treatment.

In contrast to the invariant-mass distribution of \reffi{fig:winv},
photon recombination does not produce any significant effect for the
observables $\phi,\phi'$. This is because adding a soft or collinear photon
to a fermion momentum does not change its direction significantly and,
thus, has only a small influence on the angles $\phi,\phi'$.

\begin{figure}
\setlength{\unitlength}{1cm}
\centerline{
\begin{picture}(7.7,8)
\put(-1.7,-14.5){\includegraphics{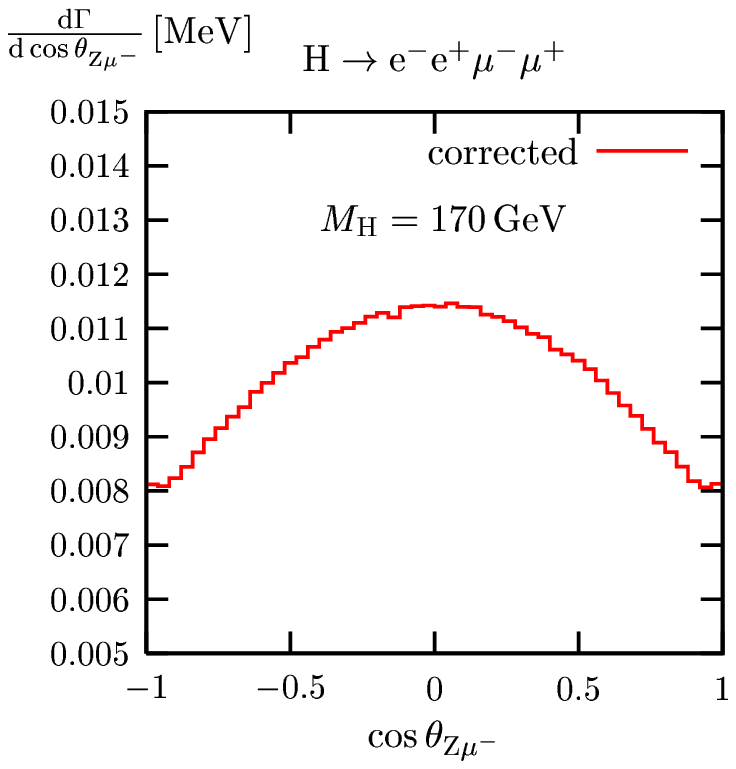}}
\end{picture}
\begin{picture}(7.5,8)
\put(-1.7,-14.5){\includegraphics{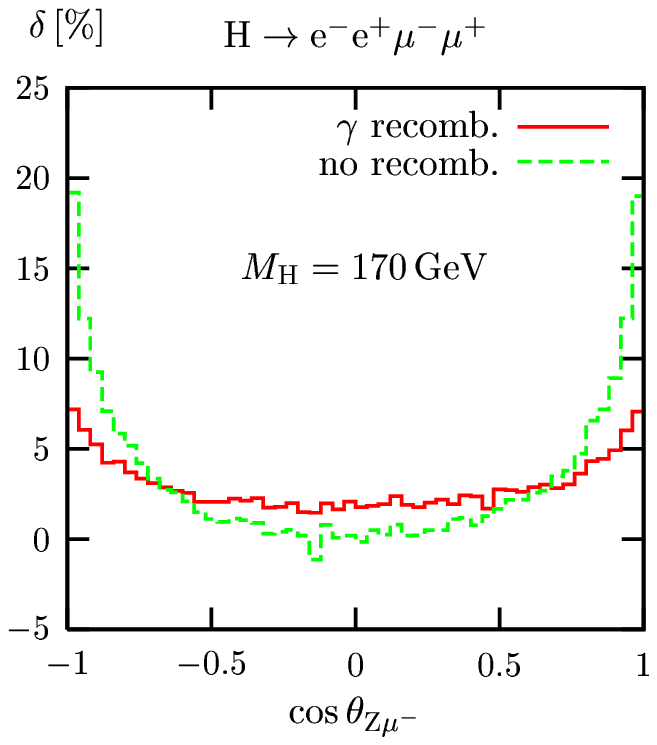}}
\end{picture} }
\centerline{
\begin{picture}(7.7,8)
\put(-1.7,-14.5){\includegraphics{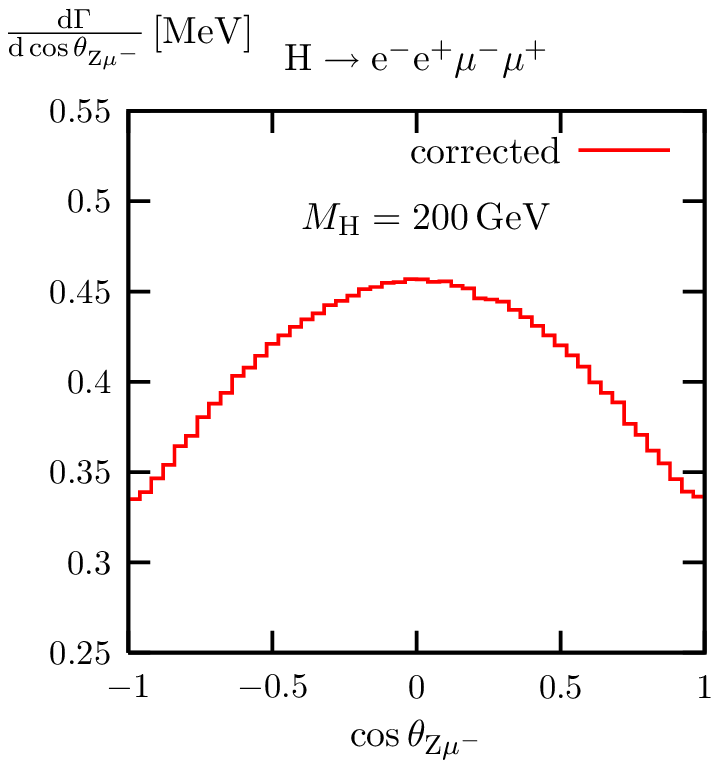}}
\end{picture}
\begin{picture}(7.5,8)
\put(-1.7,-14.5){\includegraphics{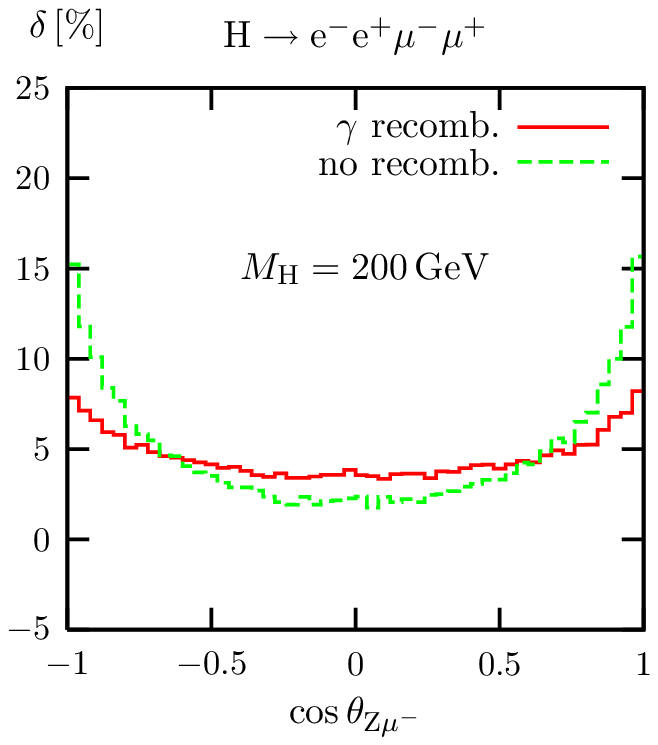}}
\end{picture} }
\caption{Distribution in the angle between the $\Pmum$ and the 
  corresponding Z~boson in the rest frame of the Z~boson (l.h.s.)\ and
  corresponding relative corrections with and without photon
  recombination (r.h.s.)\ in the decay $\PH\to\Pem\Pep\Pmum\Pmup$ for
  $\MH=170\GeV$ and $\MH=200\GeV$.}
\label{fig:th}
\end{figure}

The distribution in the decay angle of the $\Pmum$ relative to the
corresponding Z~boson in the decay $\PH\to\Pem\Pep\Pmum\Pmup$ is shown
in \reffi{fig:th}.
The angle is defined in the rest frame of the Z~boson.  Since the
Z~bosons resulting from Higgs decay are preferably longitudinally
polarized, the distribution involves a component proportional to
$\sin^2\theta_{\PZ\Pmum}$.  The relative corrections which are shown
in the plot on the r.h.s.\ reveal a strong enhancement in the forward
and backward direction if no recombination is applied.  This
enhancement is due to events where the $\Pmup$ emits a collinear
photon and has only a small energy left. Since the momentum of the
Z~boson is defined via its decay fermions, it has almost the same
momentum as the $\Pmum$. This configuration is enhanced by collinear
logarithms which are not compensated by virtual contributions.  After
applying photon recombination, the momentum of the Z~boson is defined
via the sum of the fermion and photon momenta.  Thus, the $\Pmum$ is
not necessarily collinear to the Z~boson anymore, and events are
rearranged to smaller $|\cos\theta_{\PZ\Pmum}|$ giving rise to a
flatter distribution.

Next, we consider the distribution in the angle between two fermions.
In the case of $\PH\to\PW\PW$ the angle between the charged fermions
can be used to discriminate the Higgs signal events from background
events \cite{Dittmar:1996ss}, because the fermions are emitted
preferably in the same direction. This can be understood as follows.
At leading order, the only non-vanishing helicity amplitudes for
$\PH\to\PW\PW$ are those with equal-helicity W~bosons.  Since W~bosons
only couple to left-handed particles and due to angular momentum
conservation, particles (anti-particles) are emitted preferably in the
forward direction of transverse W~bosons with negative (positive)
helicity, and anti-particles (particles) in the backward direction.
As, close to threshold, 2/3 of the W~bosons are transverse and as the
W~bosons fly in opposite directions, a particle and an anti-particle
of their decay products will be emitted preferably in the same
direction, resulting in small angles between these particles.

In the decay $\PH\to\Pne\Pep\Pmum\Pnmubar$ neither the Higgs-boson nor
the W-boson momenta can be reconstructed from the decay products.  The
distribution in the angle between the $\Pep$ and $\Pmum$ can, thus,
only be studied upon including the Higgs-production process.  If the
Higgs boson was, however, produced without transverse momentum, or if
the transverse momentum was known, the angle between $\Pep$ and
$\Pmum$ in the plane perpendicular to the beam axis could be studied
without knowledge of the production process.  We define the transverse
angle between $\Pep$ and $\Pmum$ in the frame where ${\bf
  k}_{\PH,\mathrm{T}}=0$ as
\beqar
\cos\phi_{\Pe\mu,\mathrm{T}} &=& 
\frac{{\bf k}_{2,\mathrm{T}}\cdot{\bf k}_{3,\mathrm{T}}}
{|{\bf k}_{2,\mathrm{T}}|{|\bf k}_{3,\mathrm{T}}|},
\nn \\
\sgn(\sin{\phi_{\Pe\mu,\mathrm{T}} }) 
       &=& \sgn\{{\bf e}_z\cdot({\bf k}_{2,\rT}\times{\bf k}_{3,\rT})\},
\eeqar
where ${\bf k}_{i,\mathrm{T}}$ are the transverse components of the
fermion momenta w.r.t.\ the unit vector ${\bf e}_z$, which could be
identified with the beam direction of a Higgs production process.

\begin{figure}
\setlength{\unitlength}{1cm}
\centerline{
\begin{picture}(7.7,8)
\put(-1.7,-14.5){\includegraphics{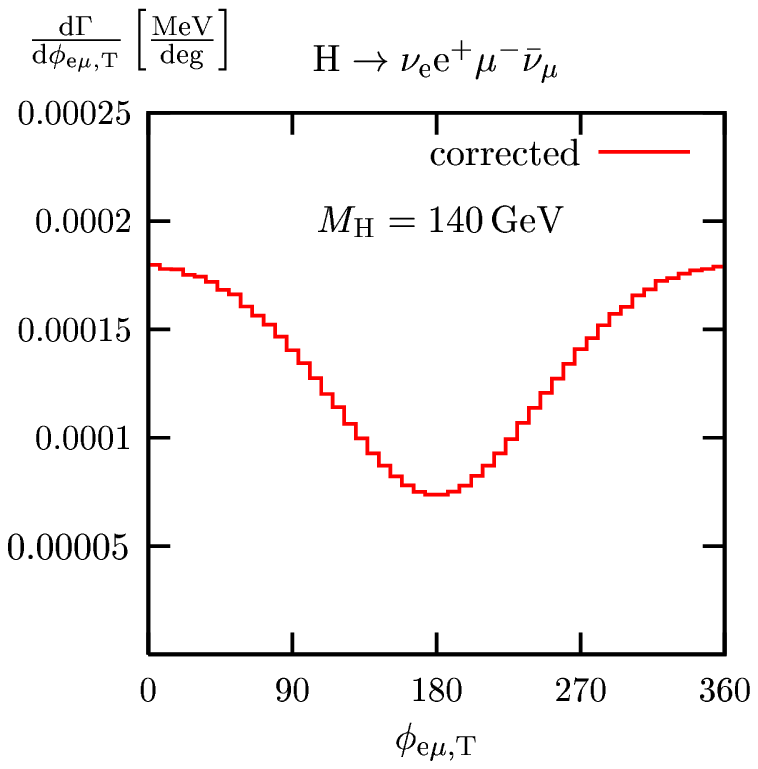}}
\end{picture}
\begin{picture}(7.5,8)
\put(-1.7,-14.5){\includegraphics{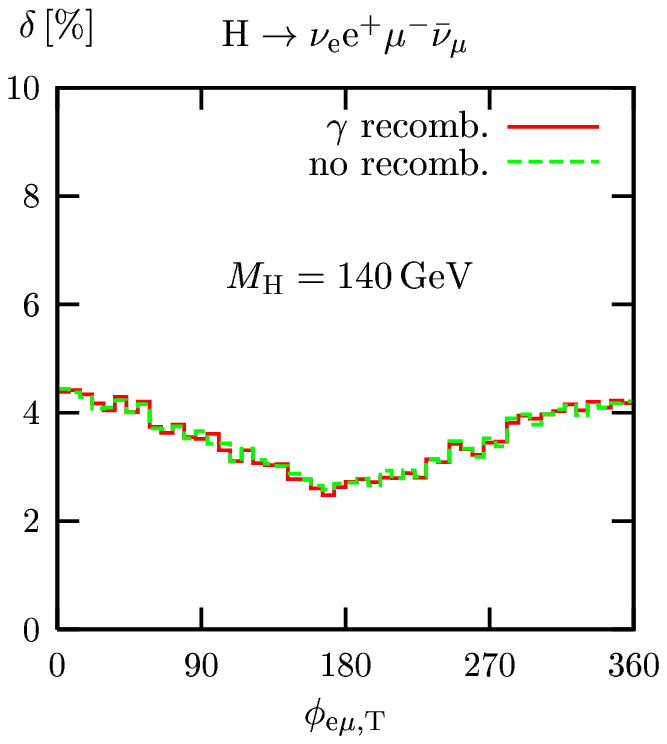}}
\end{picture} }
\centerline{
\begin{picture}(7.7,8)
\put(-1.7,-14.5){\includegraphics{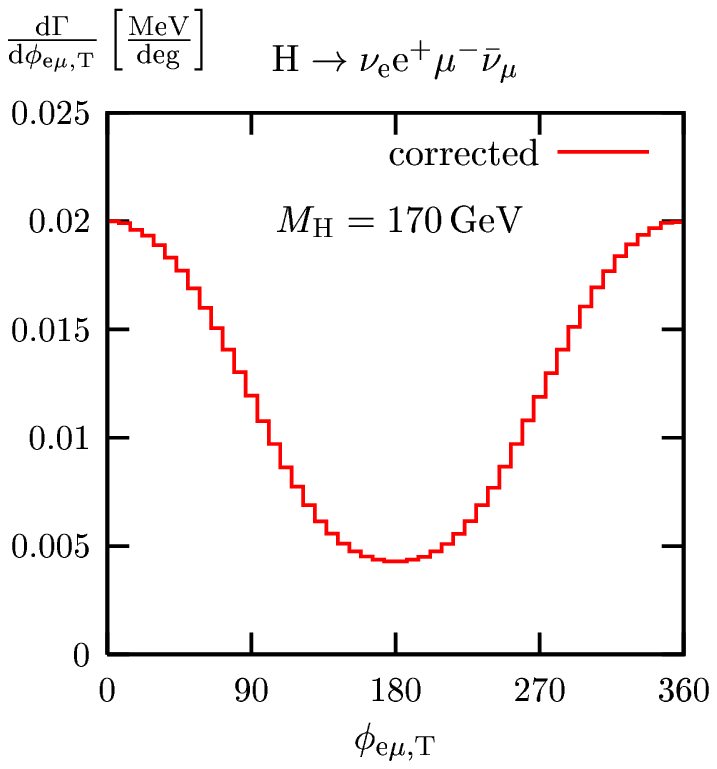}}
\end{picture}
\begin{picture}(7.5,8)
\put(-1.7,-14.5){\includegraphics{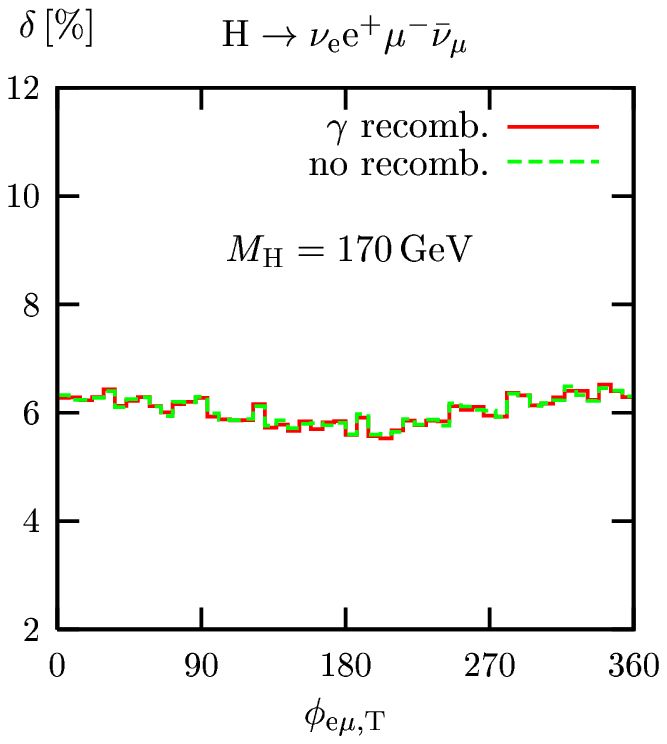}}
\end{picture} }
\caption{Distribution in the transverse angle between $\Pep$ and 
  $\Pmum$ including corrections (l.h.s.)\ and corresponding relative
  corrections (r.h.s.)\ with and without applying photon recombination
  in the decay $\PH\to\Pne\Pep\Pmum\Pnmubar$ for $\MH=140\GeV$ and
  $\MH=170\GeV$.}
\label{fig:phitr}
\end{figure}

The corresponding distribution, together with the influence of the 
corrections, is shown in \reffi{fig:phitr}. 
The enhancement for small angles, which was explained above, is
transferred to the distribution of the transverse angle
$\phi_{\Pe\mu,\mathrm{T}}$. Since the photon recombination does not
change the direction of the fermions, it does not have any visible
effect on the relative corrections.

\begin{figure}
\setlength{\unitlength}{1cm}
\centerline{
\begin{picture}(7.7,8)
\put(-1.7,-14.5){\includegraphics{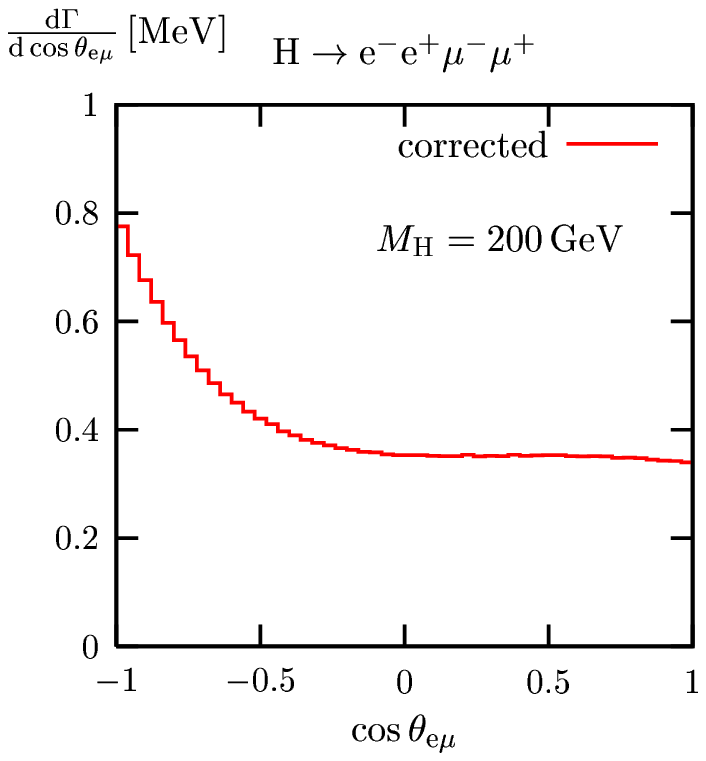}}
\end{picture}
\begin{picture}(7.5,8)
\put(-1.7,-14.5){\includegraphics{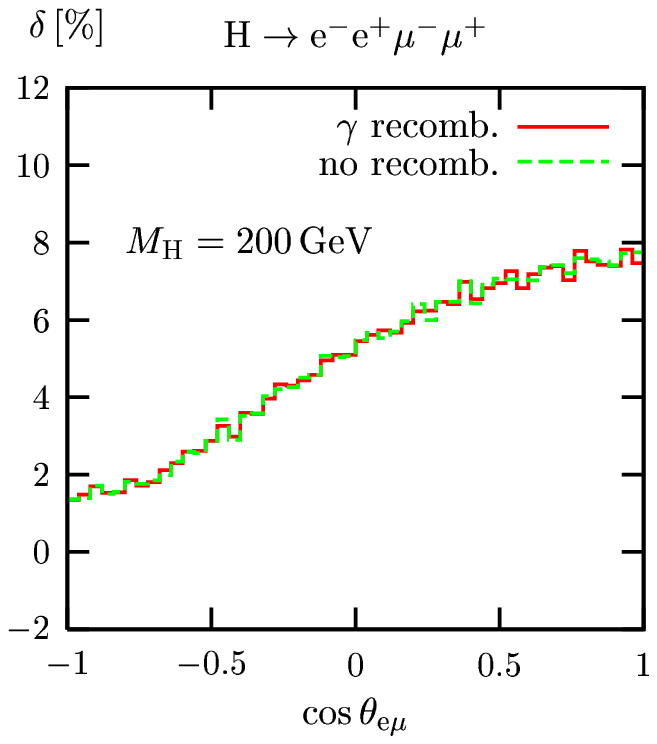}}
\end{picture} }
\caption{Distribution in the angle between $\Pem$ and 
  $\Pmum$ including corrections (l.h.s.)\ and corresponding relative
  corrections (r.h.s.)\ with and without applying photon recombination
  in the decay $\PH\to\Pem\Pep\Pmum\Pmup$ for $\MH=200\GeV$.}
\label{fig:th13}
\end{figure}

Finally, we investigate the distribution of the angle between $\Pem$
and $\Pmum$ in the decay $\PH\to\Pem\Pep\Pmum\Pmup$.  We prefer to
choose the angle between two fermions with the same charge because
this constitutes an unambiguous choice in the decay
$\PH\to\Pmum\Pmup\Pmum\Pmup$.  Figure~\ref{fig:th13} shows the
tendency that the fermions are emitted in opposite directions for the
same reason as explained above.
However, this feature is not as pronounced as in
$\PH\to\Pne\Pep\Pmum\Pnmubar$, because Z~bosons do also couple to
right-handed fermions so that one Z~boson might decay into a
left-handed fermion and the other into a right-handed fermion.
The radiative corrections tend to reduce the enhancement in forward
direction and do not depend on photon recombination.

\section{Conclusions}
\label{se:concl}

The decays of the Standard Model Higgs boson into four leptons via a
W-boson or Z-boson pair lead to experimental signatures at the LHC
that are both important for the search for the Higgs boson and for
studying its properties. To exploit this possibility a Monte Carlo
event generator for the decays $\PH\to\PW\PW/\PZ\PZ\to4\,$leptons is
needed that properly accounts for the relevant radiative corrections,
in order to achieve the necessary precision in predictions.  {\sc
  Prophecy4f} is an event generator dedicated to this task.  We have
shown first results of this generator and described the underlying
calculation.

In detail, we have presented the complete electroweak radiative
corrections of ${\cal O}(\alpha)$ to the decays $\PH\to4\,$leptons,
supplemented by corrections beyond ${\cal O}(\alpha)$ originating from
heavy-Higgs effects and final-state radiation.  The intermediate W-
and Z-boson resonances are treated in the so-called complex-mass
scheme, which fully preserves gauge invariance and does not employ any
type of expansion or on-shell approximation for the intermediate
gauge-boson resonances. Consequently, the calculation is equally valid
above, in the vicinity of, and below the WW and ZZ~thresholds.

The corrections to partial decay widths typically amount to some per
cent and increase with growing Higgs mass $\MH$, reaching about 8\% at
$\MH\sim500\GeV$. This statement, however, applies only if the
lowest-order decay widths are already evaluated with the full
off-shell effects of the intermediate W and Z~bosons, in particular
near and below the WW and ZZ~thresholds.  The on-shell (narrow-width)
approximation for the corrections is good within $0.5{-}1\%$ of the
width for Higgs masses sufficiently above the corresponding
gauge-boson pair threshold, as long as the lowest-order prediction
consistently includes the off-shell effects of the gauge bosons.  For
$\PH\to\PW\PW\to4f$ the narrow-width approximation fails by about 10\%
for Higgs masses that are only $2\GeV$ above the $\PW\PW$ threshold,
because the instability of the W~bosons significantly influences the
Coulomb singularity near threshold.  Only a calculation that keeps the
full off-shellness of the W and Z~bosons can describe the threshold
regions properly.  We have given a simple improved Born approximation
for the partial widths that reproduces the full calculation within
$\lsim2\%$ for Higgs masses below $400\GeV$. In this regime our
complete calculation has a theoretical uncertainty below 1\%. For
larger Higgs masses we expect that unknown two-loop corrections that
are enhanced by $\GF\MH^2$ deteriorate the accuracy.  Finally, for
$\MH\gsim700\GeV$ it is well known that perturbative predictions
become questionable in general.

For angular distributions, which are important in the verification of
the discrete quantum numbers of the Higgs boson, the corrections are
of the order of $5{-}10\%$ and distort the shapes.  For invariant-mass
distributions of fermion pairs, which are relevant for the
reconstruction of the gauge bosons, the situation is similar to
gauge-boson pair production processes such as
$\Pep\Pem\to\PW\PW\to4\,$fermions, i.e.\ the corrections can reach
several tens of per cent depending on the treatment of photon
radiation.

In its present version the Monte Carlo event generator {\sc Prophecy4f}
deals with fully leptonic final states, a situation most relevant for
the LHC. The generalization to semi-leptonic and hadronic final states,
including a proper description of QCD corrections, will be described
in a forthcoming publication.

\section*{Acknowledgements}

We thank M. Spira for helpful discussions about \HDECAY.

\end{document}